\documentclass[prd,superscriptaddress,nofootinbib,amsmath,amssymb,aps,11pt]{revtex4-2}

\usepackage[normalem]{ulem}
\usepackage{bm}
\usepackage{amsfonts}
\usepackage{latexsym}
\usepackage[latin1]{inputenc}
\usepackage{amsmath}
\usepackage{tabularx}
\usepackage{array}
\usepackage{palatino}
\usepackage{mathpazo}
\usepackage{graphicx} 
\usepackage{appendix}
\usepackage{booktabs}
\usepackage{dcolumn}
\usepackage[dvipsnames]{xcolor}
\usepackage{multirow}

\begin{document}

\title{Shadows of rotating hairy Kerr black holes coupled to time periodic scalar fields with non-flat target space }

\author{Galin N. Gyulchev}
	\email{gyulchev@phys.uni-sofia.bg}
	\affiliation{Department of Theoretical Physics, Faculty of Physics, Sofia University, Sofia 1164, Bulgaria}

\author{Ayush Roy}
    \email{A.Roy@soton.ac.uk}
    \affiliation{School of Mathematical Sciences and STAG Research Centre, University of Southampton,
Southampton, SO17 1BJ, United Kingdom}

\author{Lucas G. Collodel}
    \email{lucas.gardai-collodel@uni-tuebingen.de}
    \affiliation{Theoretical Astrophysics, Eberhard Karls University of T\"ubingen, T\"ubingen 72076, Germany}
    
\author{Petya G. Nedkova}
    \email{pnedkova@phys.uni-sofia.bg}
    \affiliation{Department of Theoretical Physics, Faculty of Physics, Sofia University, Sofia 1164, Bulgaria}

\author{Stoytcho S. Yazadjiev}
	\email{yazad@phys.uni-sofia.bg}	
	\affiliation{Department of Theoretical Physics, Faculty of Physics, Sofia University, Sofia 1164, Bulgaria}
	\affiliation{Institute of Mathematics and Informatics, 	Bulgarian Academy of Sciences, 	Acad. G. Bonchev St. 8, Sofia 1113, Bulgaria}

\author{Daniela D. Doneva}
	\email{daniela.doneva@uni-tuebingen.de}
	\affiliation{Theoretical Astrophysics, Eberhard Karls University of T\"ubingen, T\"ubingen 72076, Germany}

\begin{abstract}
   We study the shadows cast by rotating hairy black holes with two non-trivial time-periodic scalar fields having a non-flat Gaussian curvature of the target space spanned by the scalar fields. Such black holes are a viable alternative to the Kerr black hole, having a much more complicated geodesic structure and resulting shadows. We investigate how a nontrivial Gauss curvature alters the pictures for different amounts of scalar hair around the black holes, quantified by a normalized charge. Our results show that for high values of this charge, close to a boson star limit, chaotic shadows are observed with multiple small disconnected components for all considered Gaussian curvatures. For moderately large amounts of scalar hair and corresponding normalized charge, although the shadows still exhibit chaotic behavior, a dominant shadow component emerges, the size and shape of which are substantially influenced by the Gaussian curvature. For instance, highly chaotic shadows for flat target space, start developing a large central shadow region with the increase of the Gauss curvature even for black holes with substantially heavy scalar hair. For lower values of the normalized charge, the shadows resemble qualitatively the Kerr black hole while the Gaussian curvature has a small impact on their properties. 
\end{abstract}

\maketitle

\vspace{-0.5mm}
\section{Introduction}

Very recently the Event Horizon Telescope (EHT) Collaboration has opened up the gate to new tests of the strong field regime of gravity through observations of black hole shadows. In 2019, the EHT Collaboration captured the image of the central supermassive black hole (SMBH) in the M87 galaxy \cite{event2019first}. Three years later in 2022, the EHT Collaboration also produced an image of ${\rm Sgr} A^*$, the central supermassive black hole in the Milky Way \cite{event2022first}. This success motivated the proposition of a Next Generation EHT that will be able to take black hole snapshots with a much higher accuracy. The advance in observations pushed the theoretical development with the idea of challenging the Kerr hypothesis. Building upon the black hole images obtained by the EHT, numerous alternative models to the Kerr black hole have been proposed and examined. These include horizonless compact objects such as, naked singularities and wormholes \cite{Virbhadra:2002ju, Shaikh:2018, Sakai:2014, Shaikh:2018, Gyulchev:2018, Huang:2023, Huang:2024}, rotating regular black holes \cite{Tsukamoto_2014, PhysRevD.97.064021}, as well as beyond-Kerr black holes arising from modified theories of gravity \cite{Amarilla:2010, Amarilla:2012, Amarilla:2013, Cunha:2017, Wang:2018, Zhang:2023, Hou:2021}. While a diverse range of Kerr black hole mimickers exists, certain configurations exhibit distinct features in their shadows, leading to multi-connected or highly distorted images. Regarding this matter, an in-depth analysis of the relativistic deformability of black hole lensing images can be found in \cite{Virbhadra:2022iiy, Virbhadra:2024pru}. Examples of such scenarios comprise black holes with scalar or Proca hair \cite{herdeiro2014,herdeiro2015, Sengo:2022}, black holes in binary systems \cite{Yumoto:2012, Nitta:2011, Shipley:2016}, and black holes interacting with external matter distributions \cite{Abdolrahimi:2015, Abdolrahimi:2015a, Grover:2018}. In the present paper, we will consider further examples of scalarized black holes interacting with multiple scalar fields. These solutions can be classified according to the Gaussian curvature of the target space spanned by the scalar fields and may manifest diverse behaviors depending on its value.

Stationary asymptotically flat black hole solutions predicted by vacuum GR are hairless and completely determined by their mass and angular momentum. This also applies to the case when the Einstein equations are coupled to a single real scalar field -- it is a classical result that the black hole can not support a single scalar field hair \cite{nohair2,Heusler_1996,reviewhair}. However, the scenario changes when multiple scalar fields are considered. In the case of a complex time-dependent scalar field, or equivalently, two real scalar fields forming a flat 2-dimensional manifold (target space), it was initially discovered that black holes with hair can emerge within the perturbative regime \cite{hairysol1,hairysol2,hairysol3,hairysol4}. Shortly after that, the self-consistent non-linear solutions were generated numerically \cite{herdeiro2014,herdeiro2015}. The scalar fields in this scenario do not inherit the stationary and axisymmetric properties of the spacetime but instead have a harmonic dependence on the $t$ and $\phi$ coordinates. Even though the scalar field is time-dependent, the generalized Einstein equations, as well as the spacetime metric, are stationary. The regularity at the black hole horizon requires, though, synchronizing the angular velocity at the black hole horizon with the scalar field oscillation frequency. That is why these solutions were dubbed black holes with synchronized scalar hair. 

The black hole solutions discussed in \cite{herdeiro2014,herdeiro2015} indeed involve two scalar fields whose target space is flat, as previously mentioned. However, this represents just the simplest choice, and the geometry of the target space can be substantially more intricate \cite{Damour:1992we,horbatsch2015tensor,doneva1}. In fact, nonlinear rotating black hole solutions with synchronized hair, comprising two real scalar fields forming a non-flat manifold (non-flat target space) that is maximally symmetric, were constructed in \cite{collodel2020rotating}. It was demonstrated that the curvature of the target space can significantly alter the domain of existence and the properties of the hairy black holes.

Kerr black holes with synchronized scalar hair are ones of the very few astrophysically relevant candidates that can produce deviations from GR not only for stellar black holes but also for supermassive ones (for another interesting very recent example see \cite{Eichhorn:2023iab}). This motivated the study of their shadows that was performed in the case of a flat target space \cite{cunha2015shadows,cunha2016shadows}. The purpose of the present paper is to extend these studies to a non-flat target space, more precisely for a maximally symmetric space with curvature $\kappa$. We will systematically examine the influence of nonzero $\kappa$ on the black hole shadow and eventually explore the inverse problem -- what the back hole shadow can tell us about the target space formed by the scalar fields. 

The paper is organized as follows. In the next section, we present the scalarized black holes which we consider, describing their basic properties. In section 3 we outline the theoretical background necessary for obtaining the shadow images. Then, in section 4 we construct explicitly the shadows cast by selected solutions with positive, negative, and zero curvature of the target space. The shadow images are compared and analyzed according to the influence of the normalized charge of the solutions and the Gaussian curvature of the target space. In the last section, we present our conclusions.

\section{Kerr black holes with synchronized scalar hair -- non-flat target space geometry}
In the present paper, we focus on multiple dynamical scalar fields $\varphi^a$ minimally coupled to the Einstein gravity. The scalar fields $\varphi^a$ can be considered as generalized coordinates on an abstract $N$-dimensional Riemmanian space $({\cal E}_N,\gamma_{ab}(\varphi))$, the so-called target space. The target space metric $\gamma_{ab}(\varphi)$ should be positively defined on ${\cal E}_N$. The general action of the theory is then given by
\begin{equation}
    \label{TMST action}
    S = \frac{1}{4 \pi G} \int \sqrt{-g}\left(\frac{R}{4} - \frac{1}{2}g^{\mu \nu}\gamma_{ab}(\varphi)\partial_\mu \varphi^a \partial_\nu \varphi^b - V(\varphi) \right)\,d^4x ,
\end{equation}
where  $V(\varphi)$ is the scalar field potential. This action can be also interpreted as the vacuum action of the tensor-multi-scalar-theories of gravity (\cite{Damour:1992we,horbatsch2015tensor}). Varying it with respect to the spacetime metric and the scalar fields, we get the following field equations
\begin{eqnarray}
  \label{lucas fe metric}
    R_{\mu \nu} &=& 2 \gamma_{ab}(\varphi)\partial_\mu \varphi^a \partial_\nu \varphi^b + 2V(\varphi)g_{\mu \nu}, \\
  \label{lucas fe field}
    \square \varphi^a &=& - \gamma^a_{bc}(\varphi)g^{\mu \nu}\partial_\mu \varphi^b \partial_\nu \varphi^c + \gamma^{ab}(\varphi)\frac{\partial V(\varphi)}{\partial \varphi^b},
\end{eqnarray}
where $\square$ is the d'Alembert operator associated with the spacetime metric and $\gamma^a_{bc}(\varphi)$ are the Christoffel symbols with respect to the target space metric. 

From now on, we shall focus on two scalar fields (effectively modeling the complex scalar field originally considered in \cite{herdeiro2014}), possessing maximally symmetric target spaces $({\cal E}_2,\gamma_{ab}(\varphi))$. In this case we have globally defined coordinates, the so-called isothermal coordinates, in which the target space metric can be written in the conformally flat form 
\begin{equation}
    \gamma_{ab}(\varphi) = \Omega^2(\varphi)\delta_{ab}.
\end{equation}
Here $\delta_{ab}$ is the usual Kronecker Delta and the conformal factor $\Omega(\varphi)$ is given by
\begin{equation}
    \Omega^2(\varphi) = \frac{1}{\left(1 + \frac{\kappa}{4}\psi^2\right)^2},
\end{equation}
with $\psi^2 = \delta_{ab}\varphi^a \varphi^b$ and $\kappa$ being the Gaussian curvature of the target space.

For the scalar fields potential $V(\varphi)$ we assume the simplest standard massive potential  given by
\begin{equation}
    V(\psi) = \frac{1}{2}\mu^2 \psi^2
\end{equation}
where $\mu$ is the scalar field mass.

Since we are interested in rotating black holes, the ansatz for the stationary and axisymmetric line element is chosen to be
\begin{equation} 
\label{lucas line element}
    ds^2 = -\mathcal{N}e^{2F_0}dt^2 +e^{2F_1}\left(\frac{dr^2}{\mathcal{N}} + r^2 d\theta^2\right) +e^{2F_2}r^2 \sin^2\theta \left(d \phi - \frac{\omega}{r}dt \right)^2,
\end{equation}
where $\mathcal{N} = 1 - \frac{r_H}{r}$ ($r_H$ is the location of the horizon in these coordinates), while $F_0, F_1, F_2$, and $\omega$ are functions of $r$ and $\theta$ only.  The interested reader can consult the Appendix of \cite{herdeiro2015} for an isometry between the line element in eq. \eqref{lucas line element} and the Kerr line element in Boyer-Lindquist coordinates. 

In the present paper, we will be interested in scalar field endowed black holes. In order to violate the no-scalar-hair theorems it is not enough to consider multiple scalar fields. Additionally, one should let the fields be time-dependent \cite{reviewhair}. Thus, an ansatz for these fields, which is also consistent with the circularity of the metric (\ref{lucas line element}), is the following
\begin{equation}
\label{lucas scalar field ansatz}
    \varphi^1 = \psi(r, \theta)\cos(\omega_s t + m \phi), \quad \varphi^2 = \psi(r, \theta)\sin(\omega_s t +m \phi),
\end{equation}
where $\omega_s$ is a real parameter and $m$ is an integer. One can easily check that although $\varphi^1$ and $\varphi^2$ are time-dependent, the resulting field equations \eqref{lucas fe metric}, \eqref{lucas fe field} are stationary. More details can be found in \cite{collodel2020rotating}.

Assuming these forms of the scalar field and metric ansatze, plus the appropriate boundary conditions that are regularity at the event horizon and the axes, and asymptotic flatness at infinity, one is able to find stationary black hole solutions with nontrivial scalar hair. It is interesting to note that the regularity at the horizon leads to the condition  
\begin{equation}
    \omega|_{r=r_H} = -\frac{r_H \omega_s}{m},
    \end{equation}
which also ensures that there is no scalar flux into the black hole. Therefore, the angular velocity at the black hole horizon $\Omega_H=\omega_H/r_H$ should be equal to $\omega_s$. Thus the scalar field is synchronized with the black hole rotation. Additionally, at infinity 
\begin{equation}
    \lim_{r \to \infty} \psi \propto \frac{1}{r}exp \left(- \sqrt{\mu^2 - \omega_s^2}r\right),
\end{equation}
thereby bounding $\omega_s$ via $\omega_s^2 \leq \mu^2$. 

\begin{figure}
    \centering
    \includegraphics[width=0.95\textwidth]{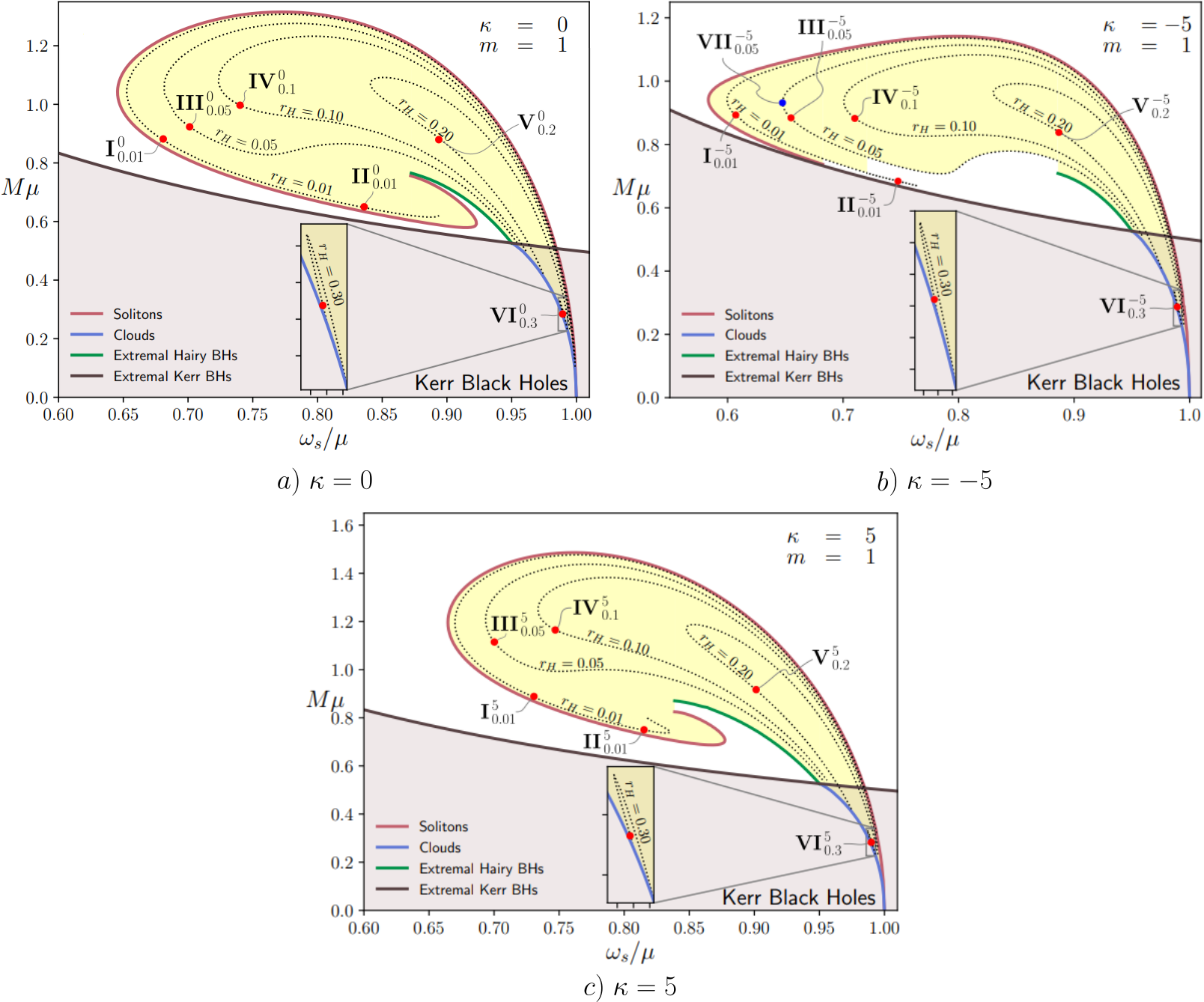}    
    \caption{\small Tracks for fixed $r_H$ in the $M-\omega_s$ plane, taken from \cite{collodel2020rotating} for $\kappa=\{0,-5,5\}$. The solutions are obtained by solving a coupled set of partial differential equations found by varying the Einstein-Hilbert action coupled to two scalar fields. The black line and below holds Kerr solutions, whereas new exotic solutions are held in the yellow bubble. It is bounded by three classes of solutions, solitonic, extremal hairy black hole, and cloud solutions, depending on the value of $r_H$ and normalized charge $q = \frac{mQ}{J}$, as explained in the text. The positions of selected solutions are indicated by red and blue dots, which, for a given $r_{H}$ and various $\kappa$, share the same $q$. Each configuration is identified by \textbf{X}$^{\,u}_{\,v}$, where the symbol \textbf{X} represents the configuration number, superscript $u$ represents $\kappa$, and the subscript $v$ denotes $r_{H}$. The physical characteristics of these selected solutions are given in Table \ref{tab_7} of the Appendix.}\label{fig:kappa}
\end{figure}

The ADM mass and angular momentum can be calculated from by the metric asymptotics
\begin{equation}
    \label{ADM M and J}
    M = \frac{1}{2} \lim_{r \to \infty} r^2 \partial_r F_0, \quad J = \frac{1}{2} \lim_{r \to \infty} r^2 \omega.
\end{equation} 
They have varying contributions from the black hole bare mass and the hair. To specify the hairiness, the normalized charge $q$ is introduced, defined as $q = \frac{mQ}{J}$, where $Q$ is the conserved Noether charge. It is commonly used to characterize Kerr black holes with synchronized scalar hair \cite{herdeiro2014} and is thus useful also for comparison reasons. Interestingly, the angular momentum of the scalar field is quantized in terms of the Noether charge $Q$, i.e, $J_{\psi} = mQ$, so we simply have $q = \frac{J_{\psi}}{J}$. This quantization is similar to the rotating boson stars \cite{Liebling:2012fv}. The $q\approx0$ solutions are the limit of scalar clouds non-backreacting the metric, whereas $q\approx1$ solutions are the boson star limit (potentially with non-flat target space metric \cite{doneva1,doneva2}).

In the present paper we will utilize the numerical solutions obtained in \cite{collodel2020rotating} 
for Gaussian curvatures $\kappa \in \{-5,0,5\}$ and fixed $m=1$. The domains of existence of hairy black holes are presented in a $M - \omega_s$ plane in Fig. \ref{fig:kappa} for different $\kappa$. The black line marks the extremal Kerr limit, with $a=M$, and thus Kerr black holes exist only below it in the grey-shaded region. The yellow region is the domain of existence of hairy Kerr black hole solutions having the following boundaries -- the red line is the solitonic (boson star) limit with $q=1$ and $r_H=0$, extremal hairy black holes are denoted with a green line (again $r_H=0$ but $q\ne 0$), while blue lines are used to depict the cloud solutions ($q=0$) \footnote{The fact that some of these lines look incomplete is due to numerical difficulties in constructing the solutions that in general become worse for more negative $\kappa$.}. A general trend is that with the decrease of $\kappa$ the domain of existence of hairy black holes becomes more deformed. Dashed lines mark sequences of solutions with constant horizon radii while red dots on these lines are the particular black hole solutions we used for building shadows. 

From the point of view of $r_{H}$, black holes with smaller values of $r_{H}$ have a larger domain of existence in $\omega_s$: They move closer to the pure solitonic ($q=1$) and extermal black hole limit, both of which have a vanishing horizon radius. Tracks denoted with an extremely small value of $r_{H}$ start and stop at the cloud line. More generally, however, each fixed track of $r_{H}$ starts at the Minkowski limit with $\omega_s/\mu=1$ and stops at the cloud line. 

\section{Calculating the black hole shadow}
\subsection{Photon Equations of Motion and Initial Conditions}

To generate the black hole's shadow through backward ray tracing, we numerically integrate photon geodesics until they either fall into the black hole or intersect the celestial sphere. The paths of these geodesics are governed by the geodesic equations, which are obtained from Hamilton's equations:
\begin{equation}\dot{x}^\mu=\frac{\partial \mathcal{H}}{\partial p_\mu}\ ,\qquad \dot{p}_\mu=-\frac{\partial \mathcal{H}}{\partial x^\mu}\ ,\label{eq_Hamilton}\end{equation}
where the overdot implies differentiation with respect to an affine parameter. The Hamiltonian is defined in terms of the contravariant components of the metric tensor, $g^{\mu\nu}$. Assuming minimal coupling between the null particles with 4--momentum $p_{\mu}$ components and the spacetime geometry, it satisfies that
\begin{equation}\label{Hamiltonian}
  \mathcal{H}\equiv\frac{1}{2}g^{\mu\nu}p_{\mu}p_{\nu}=0.
\end{equation}
Due to stationarity and axisymmetry, $\mathcal{H}$ does not depend on $t$ and $\phi$, and both $p_t$ and $p_\phi$ are conserved quantities of the geodesic motion. Assuming the asymptotically flatness of the spacetime, we can then define the integrals of motion $E$ and $L$, which are interpreted as the energy and azimuthal angular momentum of the photon as measured by an asymptotic static observer:
\begin{equation}\label{E_and_L def}
    E = - p_t = - g_{tt}\dot{t} - g_{t\phi}\dot{\phi}, \quad\quad L = p_\phi = g_{t\phi}\dot{t} + g_{\phi \phi}\dot{\phi}.
\end{equation}
Therefore, by decoupling the variables, we obtain
\begin{equation}\label{t_and_phi eq motion}
    \dot{t} = \frac{g_{\phi \phi}E + g_{t \phi}L}{g_{t \phi}^2 - g_{tt}\,g_{\phi \phi}}, \quad\quad \dot{\phi} = -\frac{g_{t \phi}E + g_{t t}L}{g_{t \phi}^2 - g_{tt}\,g_{\phi \phi}}.
\end{equation}
Taking into account that for the photon motion $\mathcal{H}=0$ is a conserved quantity along the trajectory, we can represent the Hamiltonian in the form:
\begin{equation}\label{Hamiltonian2}
       \mathcal{H}= p_{r}^{\,2} g^{rr} + p_{\theta}^{\,2} g^{\theta \theta} + V_{\rm eff}(r,\theta) = 0, 
\end{equation}
where the quantity $T\equiv p_{r}^{\,2} g^{rr} + p_{\theta}^{\,2} g^{\theta \theta}\geq0$ related to the kinetic energy is positive definite. Here, the effective potential $V_{\rm eff}(r,\theta)\leq0$ is negative definite and reveals the allowed region $(r,\theta)$ of the photon's motion. It is given by
\begin{equation}
    \label{V eff}
    V_{\rm eff}(r,\theta) = \frac{E^2g_{\phi \phi} + 2EL g_{t \phi} + L^2 g_{tt}}{g_{tt}g_{\phi \phi} - g_{t \phi}^2}=\frac{E^2g_{tt}}{(g_{tt}g_{\phi\phi}-g_{t\phi}^2)}(\eta-h_{+})(\eta-h_{-}),
\end{equation}
where $\eta=L/E$ is the impact parameter. The functions $h_{\pm}(r,\theta)$ are the photon's effective potentials defined as:
\begin{equation}\label{hpm}
    h_{\pm}=\frac{-g_{t\phi}\pm\sqrt{g_{t\phi}^2-g_{tt}g_{\phi\phi}}}{g_{tt}}.
\end{equation}
The discriminant $\mathcal{D}=g_{t\phi}^2-g_{tt}g_{\phi\phi}$ is positive outside the black hole horizon, while $g_{tt}$ is positive inside (negative outside) the ergoregion.
\begin{figure}
    \includegraphics[width=0.6\textwidth]{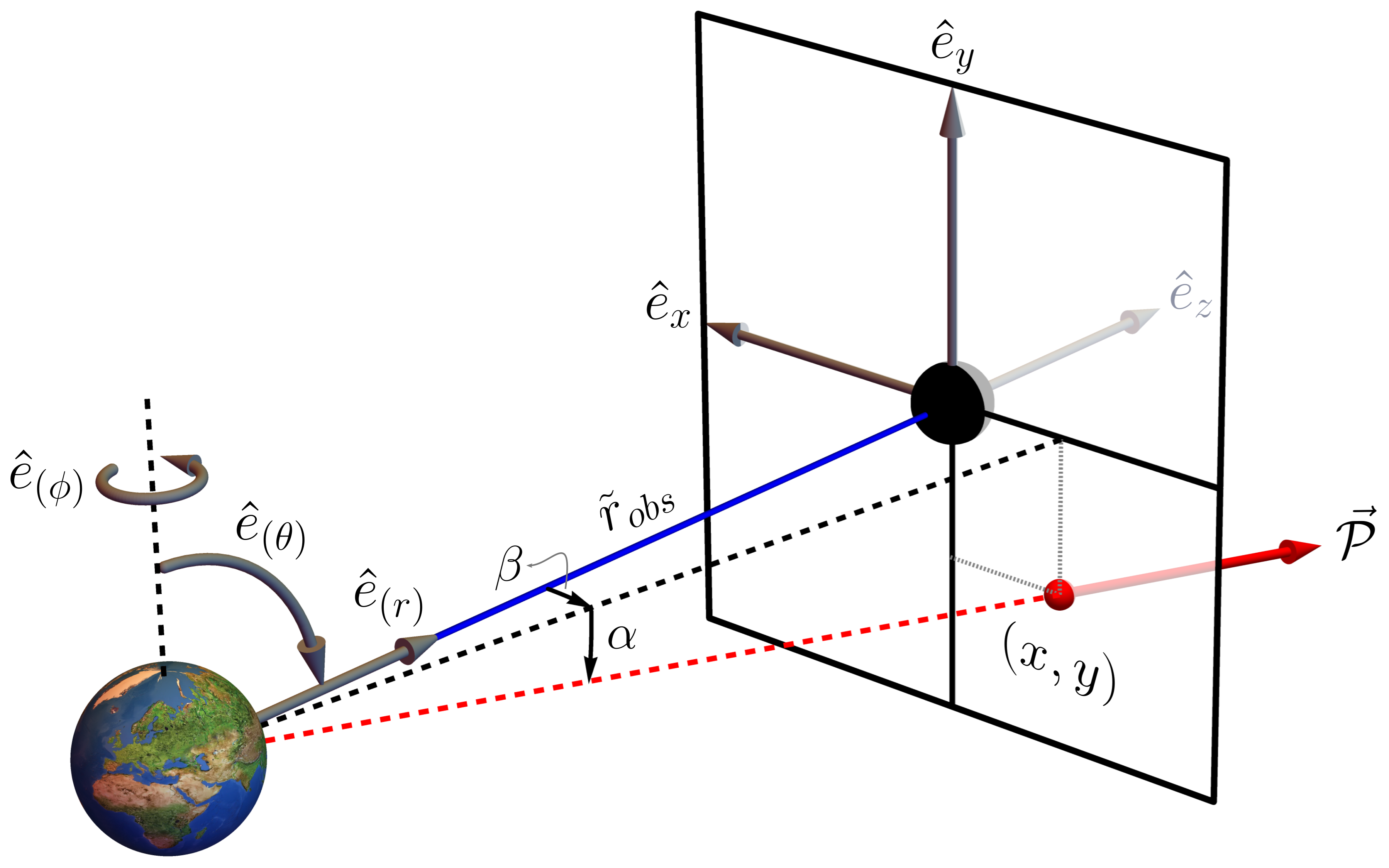}
    \caption{\small Geometric representation of the spatial momentum of the photon $\vec{\mathcal{P}}$ measured in the orthonormal basis of the observer. The observer's screen is located at a circumferential distance $\tilde{r}_{obs}$ from the black hole, whose event horizon is represented by a black sphere in the diagram. The photon impact parameters $(x, y)$ are related to the angles $\alpha$ and $\beta$, where $\alpha$ is the angle between $\vec{\mathcal{P}}$ and its projection in the $xz$-plane, and $\beta$ is the angle between this projection and the optical axis indicated in blue. The observer is oriented in such a way relative to the black hole that the radial directions coincide, i.e., $\hat{e}_{(r)} = \hat{e}_{z}$, while the basis vectors that span the image plane obey the connections $\hat{e}_{(\theta)} = -\hat{e}_{y}$ and $\hat{e}_{(\phi)} = \hat{e}_{x}$.}
	\label{fig:LensDiagram}
\end{figure} 

To begin ray tracing, we require information about a photon's initial conditions, namely $t, r, \theta, \phi, E, L, p_r$ and $p_\theta$. These are obtained by constructing a local orthonormal basis $\{\hat{e}_{(t)},\hat{e}_{(r)},\hat{e}_{(\theta)},$ $\hat{e}_{(\phi)}\}$ at the position of the observer, generally placed at $(\tilde{r}_{obs}=15M$ and $\theta_{obs}= \frac{\pi}{2})$ throughout this work. Here $\tilde{r}$ is the circumferential radius of a point at the equatorial plane and is given by:
\begin{equation}\label{rtilde}
    \tilde{r}=\frac{1}{2\pi}\int_{0}^{2\pi}d\phi\,\sqrt{g_{\phi\phi}}=r \, e^{F_{2}(r)},
\end{equation}
where $r$ is the coordinate radius. To establish the initial conditions for the photon's 4-momentum, we adhere to the formalism discussed in \cite{cunha2016shadows, lora2022osiris}. For convenience, we introduce two angles, $\alpha$ and $\beta$, to parameterize the impact parameters of a photon at the observer's location, as illustrated in Fig. \ref{fig:LensDiagram}:
\begin{equation}
    \label{shadow impact parameters}
    x = - \tilde{r}_{obs} \tan \beta , \quad y = \tilde{r}_{obs} \sin \alpha.
\end{equation}
At a chosen observer's circumferential distance $\tilde{r}_{obs}(r_{obs})$, first, we calculate the coordinate position $r_{obs}$, according to Eq. (\ref{rtilde}). Positioned at the point $(\tilde{r}_{obs}, \theta_{obs})$, the observer views a two-dimensional flat screen, $(x,y)$, passing through the center of the black hole and perpendicular to the axis between the observer and the origin. In this scenario, the observer aligns radially with the black hole, meaning that $\hat{e}_{(r)} = \hat{e}_{z}$, while the basis vectors that span the image plane are determined by the connections $\hat{e}_{(\theta)} = -\hat{e}_{y}$ and $\hat{e}_{(\phi)} = \hat{e}_{x}$. Additionally, the magnitude of the 4--momentum of the photon in the observer's frame of reference satisfies the Hamiltonian constraint 
\begin{equation}
    \mathcal{P}^2=-(\mathcal{P}^{(t)})^2+(\mathcal{P}^{(r)})^2+(\mathcal{P}^{(\theta)})^2+(\mathcal{P}^{(\phi)})^2=0.
\end{equation}    
Then, the components of the spatial momentum $\mathcal{\vec{P}}$ in terms of the angles $(\alpha,\beta)$ are given by,
\begin{equation}
    \label{shadow spatial momentum}
    \mathcal{P}^{(r)}= |\mathcal{\vec{P}}| \cos \alpha \cos \beta, \quad \mathcal{P}^{(\theta)} = |\mathcal{\vec{P}}| \sin \alpha, \quad \mathcal{P}^{(\phi)} = |\mathcal{\vec{P}}| \cos \alpha \sin \beta, 
\end{equation}
and the photon has $\mathcal{P}^{(t)}=|\mathcal{\vec{P}}|$. Since $|\mathcal{\vec{P}}|$ only determines the photon's frequency and does not alter the trajectory, it is set to unity, implying that $\mathcal{P}^{(t)} = 1$ as well.

We must now relate the components of the 4-momentum in generally curved coordinates to the quantities $\mathcal{P}^{(t)}, \mathcal{P}^{(r)}, \mathcal{P}^{(\theta)}$, and $\mathcal{P}^{(\phi)}$. This requires an understanding of how the coordinate basis $\{\partial_{t},\partial_{r},\partial_{\theta},\partial_{\phi}\}$ appear in the local orthonormal basis. The chosen basis takes the following form
\begin{equation}
    \label{shadow basis ansatz}
    \hat{e}_{(\theta)} = A^\theta \partial_\theta, \quad \hat{e}_{(\phi)} = A^\phi \partial_\phi, \quad \hat{e}_{(t)} = \zeta \partial_t + \gamma \partial_\phi, \quad \hat{e}_{(r)} = A^r \partial_r. 
\end{equation}
Now imposing the conditions $\hat{e}_{(r)} \cdot \hat{e}_{(r)} =1$, $\hat{e}_{(\theta)} \cdot \hat{e}_{(\theta)} =1$, $\hat{e}_{(\phi)} \cdot \hat{e}_{(\phi)} =1$, $\hat{e}_{(t)} \cdot \hat{e}_{(t)}=-1$ and $\hat{e}_{(t)} \cdot \hat{e}_{(\phi)}=0 $, we arrive at
\begin{equation}
    A^r = \frac{1}{\sqrt{g_{rr}}}, \quad  A^\theta = \frac{1}{\sqrt{g_{\theta \theta}}},  \quad A^\phi = \frac{1}{\sqrt{g_{\phi \phi}}}, \quad \gamma = - \zeta \frac{g_{t \phi}}{g_{\phi \phi}}, \quad \zeta = \sqrt{\frac{g_{\phi \phi}}{g_{t \phi}^2 - g_{tt}g_{\phi \phi}}}.
\end{equation}
The final form of the four basis vectors describing a Minkowski frame is as follows,
\begin{equation}
\label{shadow basis}
    \hat{e}_{(t)} = \zeta  \left(\partial_t - \frac{g _{t \phi}}{g_{\phi \phi}} \partial_\phi \right), \quad \hat{e}_{(r)} = \frac{1}{\sqrt{g_{rr}}} \partial_r, \quad \hat{e}_{(\theta)} = \frac{1}{\sqrt{g_{\theta \theta}}} \partial_\theta, \quad \hat{e}_{(\phi)} = \frac{1}{\sqrt{g_{\phi \phi}}} \partial_\phi.
\end{equation}
Using this information, we project the components of the 4-momentum into this basis as follows,
\begin{eqnarray}
&&    \mathcal{P}^{(t)} = - \hat{e}^\mu_{(t)} p_\mu = - \left(\zeta p_t + \gamma p_\phi \right) = \zeta \left(E + \frac{g_{t \phi}}{g_{\phi \phi}} L \right), \\
&& \mathcal{P}^{(r)} = \hat{e}^\mu_{(r)} p_\mu = \frac{1}{\sqrt{g_{rr}}} p_r, \\
&& \mathcal{P}^{(\theta)} = \hat{e}^\mu_{(\theta)} p_\mu = \frac{1}{\sqrt{g_{\theta \theta}}} p_\theta, \\   
&& \mathcal{P}^{(\phi)} = \hat{e}^\mu_{(\phi)} p_\mu = \frac{1}{\sqrt{g_{\phi \phi}}} L,
\end{eqnarray}
where the identification of the conserved quantities $E=-p_t$ and $L=p_\phi$ was used wherever they appear. Given the earlier presentation of the equations of motion, it is more desirable to obtain starting conditions for $E,L,p_{r}$ and $p_{\theta}$. Explicitly, inverting the above expressions yields,
\begin{eqnarray}
    && E = \frac{1+\gamma\sqrt{g_{\phi\phi}}\sin\beta\cos{\alpha}}{\zeta},  \quad  p_{r}=\sqrt{g_{rr}}\cos\beta\cos\alpha, \\
    && L=\sqrt{g_{\phi\phi}}\sin\beta\cos{\alpha},  \qquad\quad\;\; p_{\theta}=\sqrt{g_{\theta\theta}}\sin\alpha.
\end{eqnarray}

\begin{figure}
    \includegraphics[width=0.4\textwidth]{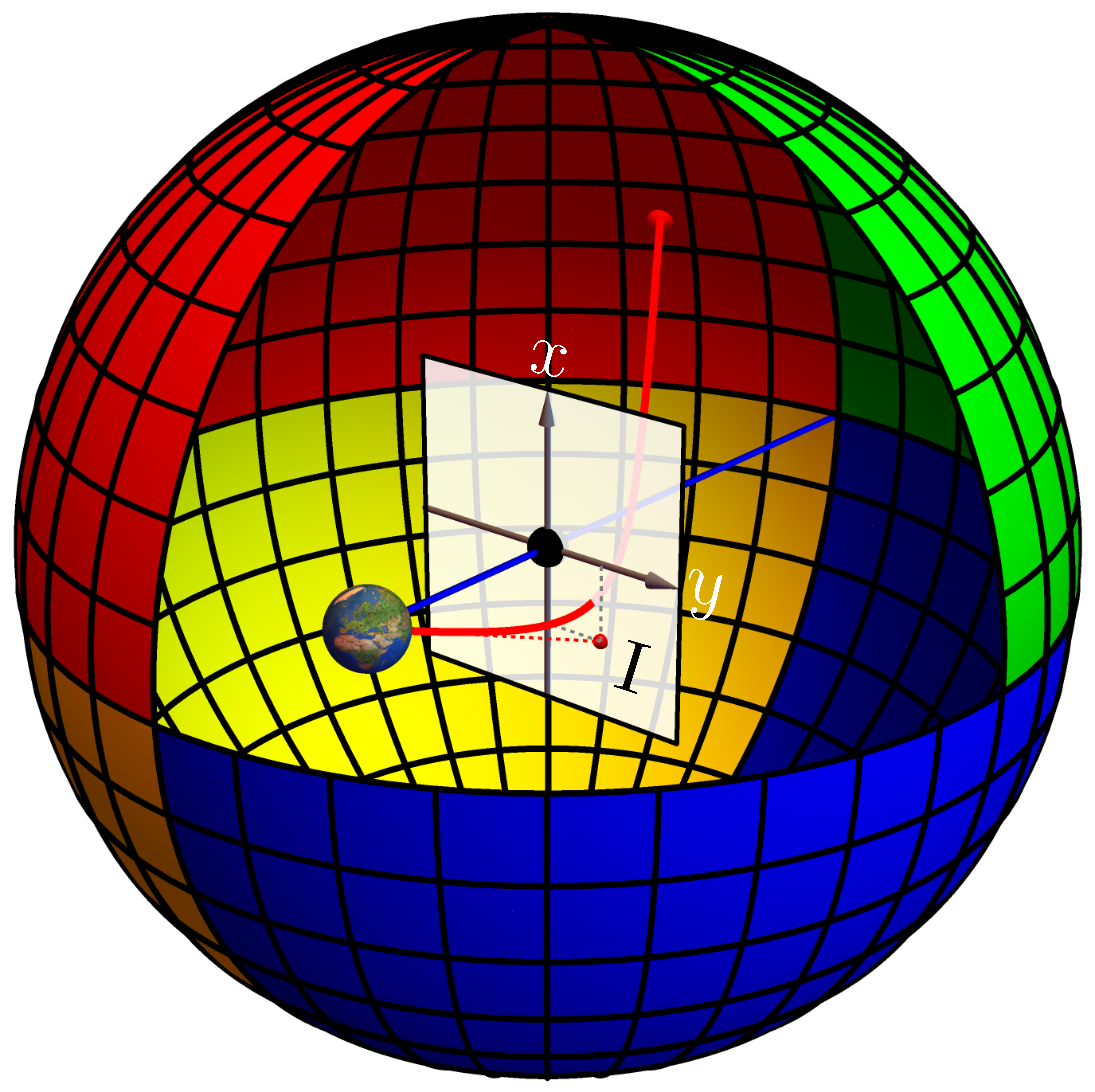}
    \caption{\small The diagram depicts the celestial sphere with half a quadrant removed, illustrating the optical elements of the gravitational system. Positioned along the blue optical axis are the observer and the black hole, the latter represented by a black sphere. The observer's screen, perpendicular to the optical axis, intersects with the black hole. A red thick line traces a specific light trajectory, characterized by impact parameters $(x,y)$, originating from Earth and terminating at a color-specified quadrant of the celestial sphere. In this geometry, the gravitational lens effect causes the observer to perceive an image $I$ with the corresponding color on the image plane. This image appears aligned with the tangent line to the light trajectory at the observer's position. The celestial sphere is divided into four quadrants, each with a distinct color designation. In the top hemisphere ($0<\theta_{cel}<\pi/2$): the green quadrant if $0<\phi_{cel}<\pi$, and the red quadrant if $\pi<\phi_{cel}<2\pi$. In the bottom hemisphere ($\pi/2<\theta_{cel}<\pi$): the blue quadrant if $0<\phi_{cel}<\pi$, and the yellow quadrant if $\pi<\phi_{cel}<2\pi$.}
	\label{fig:ColorSph}
\end{figure} 
After positioning the observer on the equatorial plane $(\theta = \pi/2)$ at a circumferential radius $\tilde{r}_{obs} = 15M$, the angles $\alpha$ and $\beta$ determine the impact parameters $(x, y)$ and initial momenta using the local orthonormal basis. With these initial conditions, photons evolve backward in time until they encounter the celestial sphere, situated at a circumferential distance $\tilde{r}_{cel}=30M$ or they hit the black hole horizon $r_{H}$. To interpret the resulting gravitational lensing patterns, we divide the celestial sphere into four quadrants, each assigned a distinct color based on the angular distribution between the polar angle $\theta_{cel}$ and the azimuthal angle $\phi_{cel}$ as demonstrated on Fig. \ref{fig:ColorSph}. Furthermore, to visualize the deformation of the characteristic patterns in the images, we introduce a grid of thin black meridian and parallel lines spaced $10^\circ$ apart in each quadrant.

\subsection{Light Rings and Ergoregions}

In this subsection, we will consider the properties of the light rings and ergoregions for the chosen configurations shown in Fig. \ref{fig:kappa}. As we will see, some of the studied configurations have stable light rings that, placed in their ergoregions, become light beam attractors, which can lead to the appearance of chaotic patterns in the black hole shadows.  

The light rings are circular null geodesics considered in the equatorial plane of symmetry, $\theta=\pi/2$, that satisfy the conditions $p_{\theta}=0$ and $p_{r}=\dot{p_{r}}=0$, from which it follows that 
\begin{equation}\label{VeffSystem}
    V_{\rm eff}=0, \quad\quad \partial_{r} V_{\rm eff}=0,
\end{equation}
at the location of the orbit. Following Eq. (\ref{V eff}), these conditions can be reduced to the next system of two algebraic equations that must be satisfied simultaneously 
\begin{eqnarray}
    && g_{\phi\phi}+2\eta g_{t\phi}+g_{tt}\eta^2=0, \label{LReq1} \\  
    && \partial_{r} g_{\phi\phi}+2\eta\partial_{r}g_{t\phi}+\eta^2\partial_{r}g_{tt}=0. \label{LReq2}
\end{eqnarray}
Solving the first of these equations for the impact parameter $\eta=L/E$ and substituting in the second, we obtain the light ring equation, which predicts the existence of a photon circular orbit with a radius $r_{LR}$
\begin{equation}\label{LReq3}
    \partial_{r} g_{\phi\phi}+2h_{\pm}\partial_{r}g_{t\phi}+h_{\pm}^2\partial_{r}g_{tt}=0,
\end{equation}
where the functions $h_{\pm}$ are defined via Eq. (\ref{hpm}). Moreover, the radial condition for the existence of stable (unstable) light rings imposes $\partial^{2}V_{\rm eff}>0$, ($\partial^{2}V_{\rm eff}<0$), which is reduced to the restriction
\begin{equation}\label{LReq4}
       \partial^2_{r} g_{\phi\phi}+2h_{\pm}\partial^2_{r}g_{t\phi}+h_{\pm}^2\partial^2_{r}g_{tt} 
        \begin{cases}
            <0, & \text{if light ring is stable.} \\
            >0, & \text{otherwise.}
        \end{cases}
    \end{equation}
Besides, taking into account that $\eta=h_{\pm}$ on the circular orbit and $h_{+}\neq h_{-}$ outside the horizon, after straightforward calculations, one can show that the conditions for the effective potential (\ref{VeffSystem}) are reduced to an analogical light ring equation
\begin{equation}\label{LReq6}
    \partial_{r}h_{\pm}=0.
\end{equation}
The solutions to that equation predict stable (unstable) light rings if the radial condition $\pm\partial^{2}_{r}h_{\pm}>0$, ($\pm\partial^{2}_{r}h_{\pm}<0$) is satisfied. Moreover, the normalized timelike Killing vector field at
infinity, denoted as $\partial_{t}$, becomes null over the surface $g_{tt}=0$, defining the ergoregions \cite{PhysRevD.89.124018}. In that special case, in the limit $g_{tt}\rightarrow 0$, one of the functions $h_{\pm}$ diverges, and the other converges to $-g_{\phi\phi}/2g_{t\phi}$. Generally, outside the ergoregions $g_{tt}<0$, while passing within the ergoregions $g_{tt}>0$. 

Light rings can also be classified in terms of their direction of rotation. Taking into account Eqs. (\ref{t_and_phi eq motion}), as well as that Eq. (\ref{LReq1}) is satisfied on the light ring, one can show that the angular velocity, $\Omega = d\phi/dt$, of the photons moving on circular orbits is connected to the impact parameter via the relation
\begin{equation}\label{Omega}
    \Omega=\frac{1}{\eta}.
\end{equation}
Hence, it follows that the light ring's rotational direction is given by the sign of the impact parameter $\eta$ for a static observer at spatial infinity. In general, the orbital angular frequency of rotating photons at the light rings is given by 
\begin{equation}\label{Omega_pm}
    \Omega_{\pm}=\frac{-\partial_{r}g_{t\phi}\pm\sqrt{\partial_{r}g_{t\phi}^2-\partial_{r}g_{tt}\partial_{r}g_{\phi\phi}}}{\partial_{r}g_{\phi\phi}},
\end{equation}
where the above expressions are evaluated at the location of the light ring. In the above equation $\Omega_{+}$ is the angular frequency of co-rotating photons, and $\Omega_{-}$ is the angular frequency of the counter-rotating photons. 
    
\subsection{Numerical Ray Tracing Methods}

To compute the cast shadows of rotating tensor-multiscalar black holes with scalar hair, a program code was developed using the Wolfram Language Mathematica. The code uses built-in interpolation and differential solver routines to integrate the Hamiltonian equations of motion numerically. For the numerical implementation of the equations, the metric functions $F_{0}$, $F_{1}$, $F_{2}$, and $\omega$ were interpolated using a two-dimensional cubic spline interpolation concerning the variables $r$ and $\theta$, which was set as the default interpolation procedure in the programming language. In the Hamiltonian equations, the interpolated functions were substituted into the analytical expressions for the metric functions and their derivatives concerning $r$ and $\theta$. For the numerical solution of the system of differential equations, we utilize time integration methods inherent to Mathematica, such as the Adams/BDF multi-step method with automatic step size control.

Additional information is required to generate the shadows' image on the observer's screen. This includes the observer's position, the two-dimensional width of the field of view, the image resolution, and the radius of the celestial sphere where the rays, scattered by the black hole, will complete their trajectory. 
 
In this context, without loss of generality, we choose the observer to be located on the equatorial plane at a specific circumferential distance from the black hole, such that $(t_{obs}, r_{obs}, \theta_{obs}, \phi_{obs}) = (0, r_{obs}(\tilde{r}_{obs}), \pi/2, 0)$. By selecting the circumferential distance as $\tilde{r}_{obs}=15M$, we use Eq. (\ref{rtilde}) to numerically determine the radial coordinate of the observer, $r_{obs}$. Simultaneously, we set the circumferential distance $\tilde{r}_{cel}=30 M$ as the second boundary condition for any geodesic reaching the celestial sphere. After performing the ray tracing procedure, we apply an equiangular projection that directly maps the angles $\theta_{cel}$ and $\phi_{cel}$, labeling the photon scattered by the black hole onto the celestial sphere with the observed angles $(\beta, \alpha)$ onto the coordinate axes $(x, y)$. To implement the projection, we establish the desired field of view, which specifies the angles between the optical axis and the image boundaries. Subsequently, by determining the desired image resolution, corresponding to $1024\times1024$ photon trajectories, we found the required step size for uniform grid formation across the entire field of view.

Increasing the geodesic flux density is crucial to generating a high-resolution image of the shadow, revealing more precisely the complexity of the chaotic patterns. To significantly diminish the integration time for the entire field of view, we first take advantage of the reflection symmetry inherent in the shadow image as perceived by an equatorial observer. This involves integrating only half of the shadow--specifically, the geodesics either above or below the equatorial plane of symmetry. Subsequently, exploiting the reflection symmetry of the examined solution enables us to reconstruct the complete shadow image, filled with its intrinsic colours denoting the geodesics scattered to different parts of the celestial sphere.

On the other hand, since modern workstations contain multiple computing cores, we utilize the built-in FinestGrained method to break down the overall computation into the minor possible subunits, whose evaluations take different amounts of time. Employing this approach facilitates
the optimization of the integration time, a notably time-consuming process for scalarized solutions with stable light rings. The geometry of these solutions contributes to the chaotic behavior of geodesics near the event horizon. This is illustrated by the shadow image and the heat map of the time delay function in Fig. \ref{fig:2}. This function is defined as the variation of the time required for a photon geodesic to travel from a particular pixel on the observer's screen to a corresponding point on the celestial sphere, measured in units of $\mu^{-1}$. The heat map provides valuable information about the distribution of time-delay function values on the observer's screen. It is particularly useful in diagnosing the light ring system since photon trajectories that come close to the light rings take a significant amount of time to return to spatial infinity. For a more detailed discussion of the time delay heat map in various scenarios, such as rotating boson stars and hairy black hole solutions, please refer to \cite{PhysRevD.94.104023}.

\section{Results}

In this section, we will examine the shadow images for seven groups of selected configurations, \textbf{I}--\textbf{VII}, all highlighted in Fig. \ref{fig:kappa}. Each configuration is identified by \textbf{X}$^{\,u}_{\,v}$, where the symbol \textbf{X} represents the configuration number, superscript $u$ represents the value of the Gaussian curvature $\kappa$, and the subscript $v$ denotes the value of the black hole horizon $r_{H}$. Each group contains three configurations for a unique normalized charge, certain black hole horizon, and various Gaussian curvatures $\kappa\in\{-5, 0, 5\}$. Generally, the higher values of $\kappa$ reduce the chaotic patterns in the shadows of black holes with different horizons, affecting the radii and rotational direction of the light rings (LR) and the equatorial radii of the ergoregions (ER). Henceforth, in the paper, we will utilize the dimensionless compactified radial coordinate $R\in[0, \, 1]$ defined as follows:
\begin{equation}
    R=\frac{\tilde{R}}{1+\tilde{R}}, \quad \mbox{with} \quad \tilde{R}=\sqrt{r^2-r_{H}^2}. 
\end{equation}

In the following subsection, we explore a particular configuration, \textbf{VII$^{\,-5}_{\,0.05}$}, which allows us to observe not only a complex structure of ergoregions and a system of multiple light rings but also a variety of non-simply connected shadow images, accompanied by the chaotic behavior of scattered photons.

\subsection{Photon Potential, Ergoregions and Light Rings}

In this section, we will focus on a specific configuration distinguished by one of the most intricate shadows. Beyond the appearance of numerous non-simply connected shadows with various shapes and regions highly saturated with chaotically scattered orbits, there lies a complex structure of light rings and multiple ergoregions. A notable aspect of this particular solution is the presence of a stable light ring characterized by torus topology situated within one of the ergoregions. To explore the mechanisms behind the formation of these unusual shadows, in this subsection, we will present the contour plots of the $h_{+}$ and $h_{-}$ photon potentials, the shadow image, and the photon's time delay heat map for the configuration \textbf{VII$^{\,-5}_{\,0.05}$}, with Gaussian curvature $\kappa=-5$, a black hole horizon of $r_{H}=0.05$ and a normalized charge $q\simeq0.996$. 

\begin{figure}
    \includegraphics[width=0.99\textwidth]{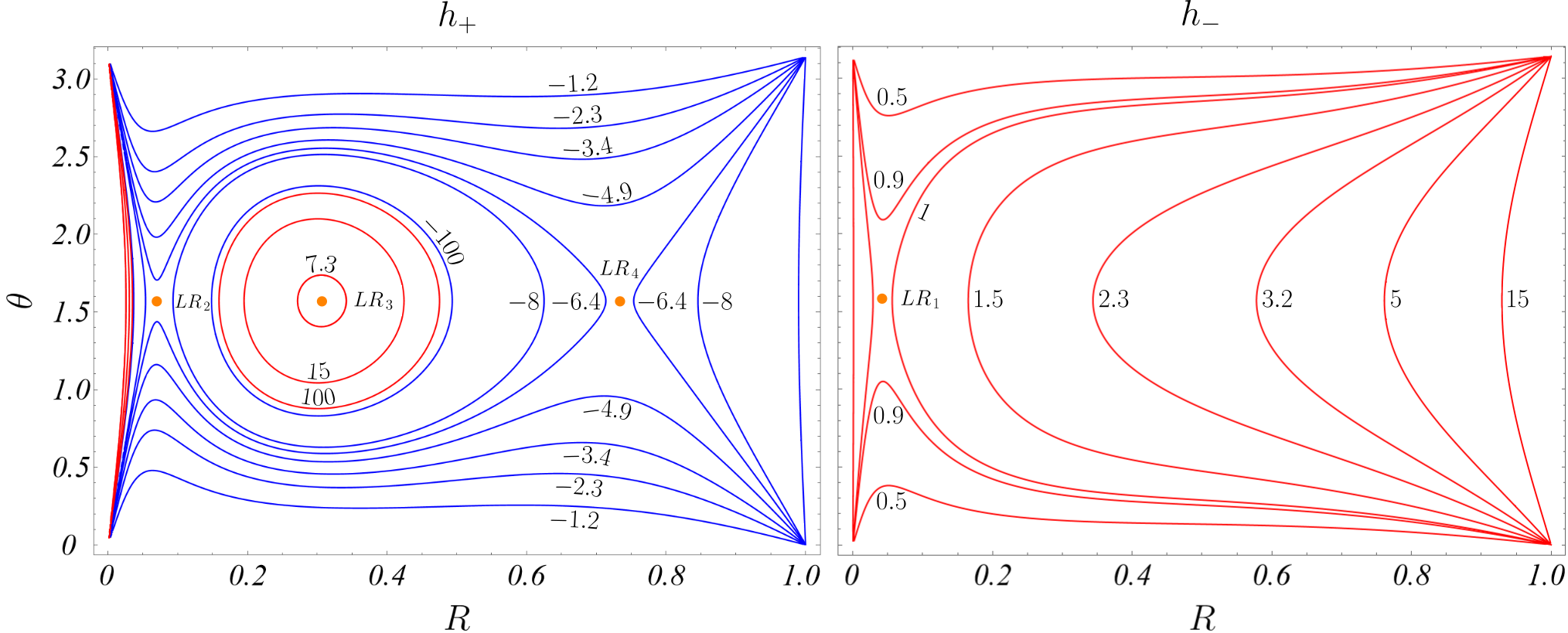}
    \caption{\small Contour plots of the two effective potentials $h_{+}$ (\textit{left panel}) and $h_{-}$ (\textit{right panel}) for configurations \textbf{VII}$^{\,-5}_{0.01}$ at Gaussian curvature $\kappa=-5$. The configuration contains two disconnected ergoregions pointed out with red contours on the upper panel, the first among which is located near the horizon. Each of the potentials $h_{+}$ and $h_{+}$ possess a saddle point between both ergoregions, allowing two unstable, closely spaced light rings to form. In the center of the second ergoregion, the potential $h_{+}$ possesses a local minimum, corresponding to a third stable light ring. Beyond the second ergoregion, the function $h_{+}$ has a second saddle point, which causes the formation of a fourth outermost unstable light ring. The orange dots indicate the locations of the four light rings.}
	\label{fig:2.1}
\end{figure} 

Fig. \ref{fig:2.1} exhibits the effective potentials $h_{+}$ and $h_{-}$ for the spacetime configuration \textbf{VII}$^{\,-5}_{\,0.05}$, and Table \ref{tab_7} in the Appendix presents the corresponding physical quantities of the selected solution. The contour lines of the function $h_{+}$ reveal a singular behavior of the potential at the boundary of two detached ergoregions, the first located near the black hole's event horizon ($R_{H}=0$) in the equatorial domain $R_{ER}\in[0, \, 0.036]$. The second ergoregion extending in the interval $R_{ER}\in[0.154, \, 0.484]$, possesses toroidal topology and contains a global minimum corresponding to the existence of an equatorial stable light ring for $R_{LR}\simeq0.307$, rotating in the same direction as the black hole. The presence of two saddle points reveals the existence of two equatorial unstable light rings, the first located between the two ergoregions for $R_{LR}\simeq0.070$ and the second formed beyond the ergotorus for $R_{LR}\simeq0.733$. Both rings correspond to photons, circling the black hole in the opposite spinning direction as the black hole.

Fig. \ref{fig:2.1}, bottom panel, exhibits $h_{-}$ contour lines, which demonstrate the existence of a saddle point, corresponding to an equatorial unstable light ring situated at $R_{LR}\simeq0.041$ outside the inner ergoregion. The sign of $\eta$ indicates that the photons in the light ring are circling the black hole in the same spinning direction as the black hole. 

The sign of $\eta$, $d\phi/dt$ and $g_{tt}$ as well as the corresponding values of the dimensionless coordinate $R$, pointing the positions of the light rings and the equatorial domain of existence of the ergoregions are organized in the following table.

\begin{center}
\setlength{\tabcolsep}{4.4pt} 
\renewcommand{\arraystretch}{1} 
\begin{tabular}{c*{10}{c}cc}
\hline
\hline
Configuration & Fig. & Ergoregions & $R_{ER}$ & LR & $R_{LR}$ & stability & $\eta$ & $g_{tt}$ & $d\varphi/dt$ & Chaos \\
\hline
\multirow{4}{*}{\textbf{VII}$^{\,-5}_{\,0.05}$} & \multirow{4}{*}{\ref{fig:2}} &                          &                           & $h_-$ & 0.041 & unstable & $+$ & $-$ & $+$ & \multirow{4}{*}{Yes} \\
                        &                              & 1 ER & $[0, \; 0.036]$     & $h_+$ & 0.070 & unstable & $-$ & $-$ & $-$ & \\
                        &                              & 2 ER & $[0.154, \;0.484]$ & $h_+$ & 0.307 & stable & $+$ & $+$ & $+$  & \\
                        &                              &                          &                          & $h_+$ & 0.733 & unstable & $-$ & $-$ & $-$ & \\
\hline\hline		
\end{tabular}
\end{center}

\begin{figure}[t!]
    \includegraphics[width=0.85\textwidth]{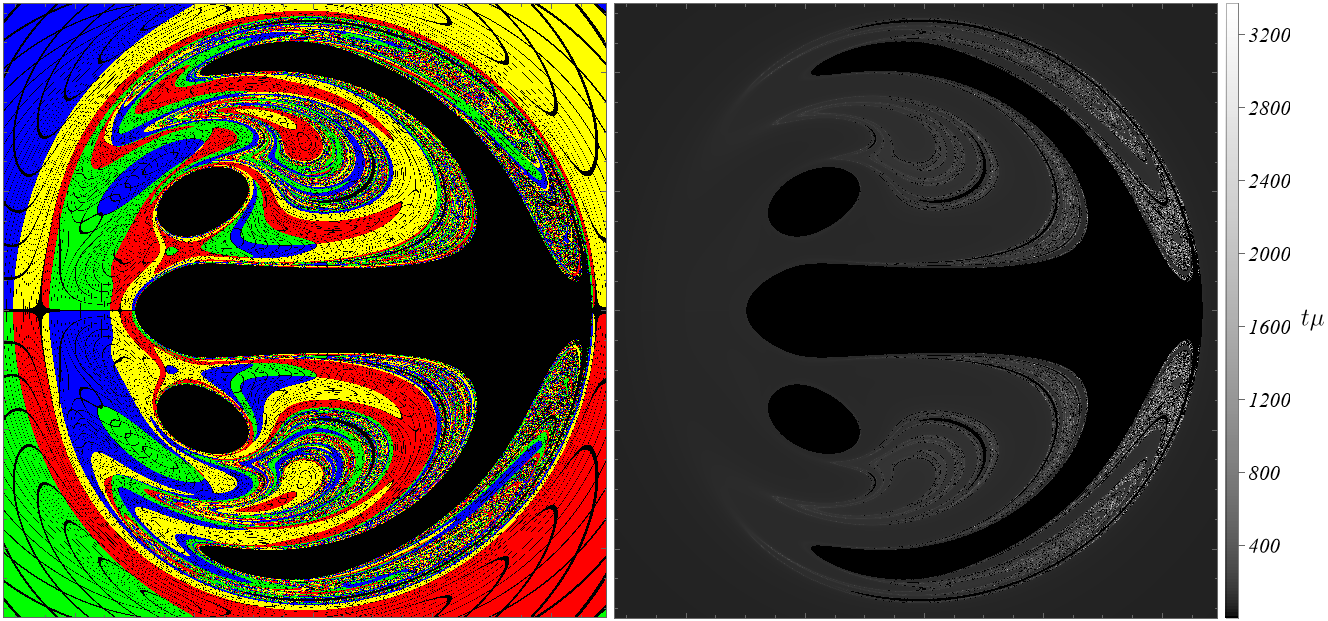}
	\caption{\small Zoom time delay heat map associated with the scattering orbits (right panel) and lensed image (left panel) for configuration \textbf{VII}$^{\,-5}_{\,0.05}$, with $\kappa=-5$, $r_H=0.05$, $\omega_{s}/\mu=0.648538$,  $M\mu=0.915671$ and $q=0.996185$. The regions corresponding to shadow points are shown in black. To clarify the interpretation of the color image, a color legend is presented in Fig. \ref{fig:ColorSph}, and a grid is introduced to emphasize the deformation of the images. Detailed physical quantities of the solution are provided in Table \ref{tab_7} in Appendix A.}
	\label{fig:2}
\end{figure}

A correlation has been observed between the chaotic patterns found in the shadow's image and the characteristics of the corresponding geodesic motion \cite{PhysRevD.94.104023}. In Fig. \ref{fig:2}, on the right, you can see the black hole shadow alongside the time delay heat function, which is depicted in units of the reciprocal values of the scalar field mass, $\mu^{-1}$. The photon's variation in the coordinate time, $t$, illustrates the expected sensitivity in mapping between the photon's coordinates $(x, y)$ in the image plane and the corresponding arrival point on the celestial sphere or the black hole horizon. The heat map in this case is very informative, as it reveals a strong correlation between the shape of the regions with bright pixels with a significant time delay and the chaotic patterns created by the scattered photons on the shadow image. This direct relationship indicates that some photon bundle are sensitive to the subset of their impact parameters, in which photons are propagated near the light rings. 

This sensitivity arises due to the effective potential $h_{+}$, which permits the existence of quasi-bound orbits for specific impact parameters $\eta$. Some of these orbits are situated in a pocket with a narrow throat containing an unstable light ring ($LR_{4}$), enabling photons to enter the pocket. Subsequently, the photons traverse the ergotorus, housing the stable light ring ($LR_{3}$), near which the orbits undergo multiple radial turning points before leaving the pocket and reaching the celestial sphere. Since the number of radial turning points depends on the frequency of changes in the sign of $\dot{p}_{r}$ during the photon's motion along the light ray's trajectory, this can be interpreted as a deviation from Kerr spacetime, where null geodesics possess at most one turning point. All trajectories of this class, semi-trapped in a pocket, take significantly longer to escape, resulting in a more significant time delay. These light beams are depicted as bright dots on the time delay heat map shown in Fig. \ref{fig:2} on the right. 

On the contrary, for specific values of the impact parameter $\eta$, the analysis of the effective potential $h_{+}$ reveals the possible appearance of a second inner throat in the pocket. This configuration allows photons to approach the unstable light ring $LR_{2}$. Due to the radial instability of this light ring, for some impact parameters, photons may either retrace their path back into the pocket or proceed towards the fourth unstable light ring, $LR_{1}$, located near the inner ergoregion, as the potential $h_{-}$ shows. Ultimately, these photons may fall into the black hole, especially with small perturbations in the impact parameter. The configuration of two throats in the effective potential may give rise to the formation of multiple disconnected shadows resembling black hole shadows reported in \cite{cunha2015shadows}.

\subsection{Black hole shadows and influence of the Gaussian curvature}

To illustrate the impact of different Gaussian curvatures on the light ring's formation, the ergoregions of the rotating solutions, and the visual features of their shadows in the following figures, we expose the shadows of the selected configurations, emphasizing the influence of this parameter on the observed phenomena. We have limited ourselves to the solutions already generated in \cite{collodel2020rotating}, which constitute sequences of black holes with fixed horizon radii. That is why the solutions are grouped with respect to $r_H$ while the normalized charge $q$ is also kept as similar in value as possible in each group.

Given the significant variation in the shape, size, and characteristic pattern of the shadow within the domain of existence of hairy Kerr black holes, as depicted in Fig. 1, we will perform a detailed survey of the 18 most interesting cases. The physical characteristics of these cases are listed in detail in Table I of the Appendix. Our focus will predominantly be on models close to the limiting red curve in that figure, marking the boson star limit. These cases exhibit the most peculiar characteristics and the appearance of chaotic regions in the shadow image. Below, the discussion is grouped according to the horizon radius and normalized charge of the explored solutions, which, roughly speaking, indicates how far away the black hole solutions are from the boson star limit.

\subsubsection{Model I, $r_H=0.01$, $q\simeq0.999$: Black hole image domited by chaotic regions.}

To study the impact of the normalized charge, $q$, on shadow formation, ergoregions, and the system of light rings, we focus on configurations \textbf{I}$^{\,\kappa}_{\,0.01}$, with $\kappa\in\{-5,0,5\}$ (highlighted in Fig. 1). All these configurations exhibit relatively higher values of the normalized charge, approximately $q\simeq0.999$. The corresponding physical parameters for these configurations are detailed in Table \ref{tab_7} in Appendix A.

Similar to the solution \textbf{VII}$^{\,-5}_{\,0.05}$, the selected configurations feature a system of four light rings, two ergoregions, and distinct shadow images saturated with multiple chaotic patterns. Among these light rings, there is always one that remains stable, contributing to the observation of numerous chaotic regions in shadow images. The peculiarity of these configurations lies in their ability to generate multiple miniature, highly elongated, simply-connected shadows. Furthermore, even though shadows exist for negative $\kappa$, the size of the shadow is marginal, while this size increases with increasing $\kappa$.

The distinctive features of the ergoregions and light rings for the analyzed configurations are presented in the following table.

\begin{center}
\setlength{\tabcolsep}{4.5pt} 
\renewcommand{\arraystretch}{1} 
\begin{tabular}{c*{10}{c}cc}
\hline
\hline
Configuration & Fig. & Ergoregions & $R_{ER}$ & LR & $R_{LR}$ & stability & $\eta$ & $g_{tt}$ & $d\varphi/dt$ & Chaos\\
\hline
\multirow{4}{*}{\textbf{I}$^{\,-5}_{\,0.01}$} & \multirow{4}{*}{\ref{fig:4}} &                          &                           & $h_-$ & 0.010 & unstable & $+$ & $-$ & $+$  & \multirow{4}{*}{Yes} \\
                        &                              & 1 ER & $[0, \; 0.004]$     & $h_+$ & 0.012 & unstable & $-$ & $-$ & $-$ & \\
                        &                              & 2 ER & $[0.045, \; 0.474]$ & $h_+$ & 0.186 & stable & $+$ & $+$ & $+$ & \\
                        &                              &                          &                          & $h_+$ & 0.727 & unstable & $-$ & $-$ & $-$ & \\
\hline
\multirow{4}{*}{\textbf{I}$^{\,0}_{\,0.01}$} & \multirow{4}{*}{\ref{fig:4}} &                          &                           & $h_-$ & 0.010 & unstable & $+$ & $-$ & $+$ & \multirow{4}{*}{Yes} \\
                        &                              & 1 ER & $[0, \; 0.003]$     & $h_+$ & 0.012 & unstable & $-$ & $-$ & $-$ & \\
                        &                              & 2 ER & $[0.063, \; 0.462]$ & $h_+$ & 0.196 & stable & $+$ & $+$ & $+$ & \\
                        &                              &                          &                          & $h_+$ & 0.722 & unstable & $-$ & $-$ & $-$ & \\
\hline
\multirow{4}{*}{\textbf{I}$^{\,5}_{\,0.01}$} & \multirow{4}{*}{\ref{fig:4}} &                          &                           & $h_-$ & 0.009 & unstable & $+$ & $-$ & $+$ & \multirow{4}{*}{Yes} \\
                        &                              & 1 ER & $[0, \; 0.005]$     & $h_+$ & 0.013 & unstable & $-$ & $-$ & $-$ & \\
                        &                              & 2 ER & $[0.043, \; 0.445]$ & $h_+$ & 0.155 & stable & $+$ & $+$ & $+$ & \\
                        &                              &                          &                          & $h_+$ & 0.718 & unstable & $-$ & $-$ & $-$ & \\
\hline\hline		
\end{tabular}
\end{center}

\begin{figure}
    \includegraphics[width=0.33\textwidth]{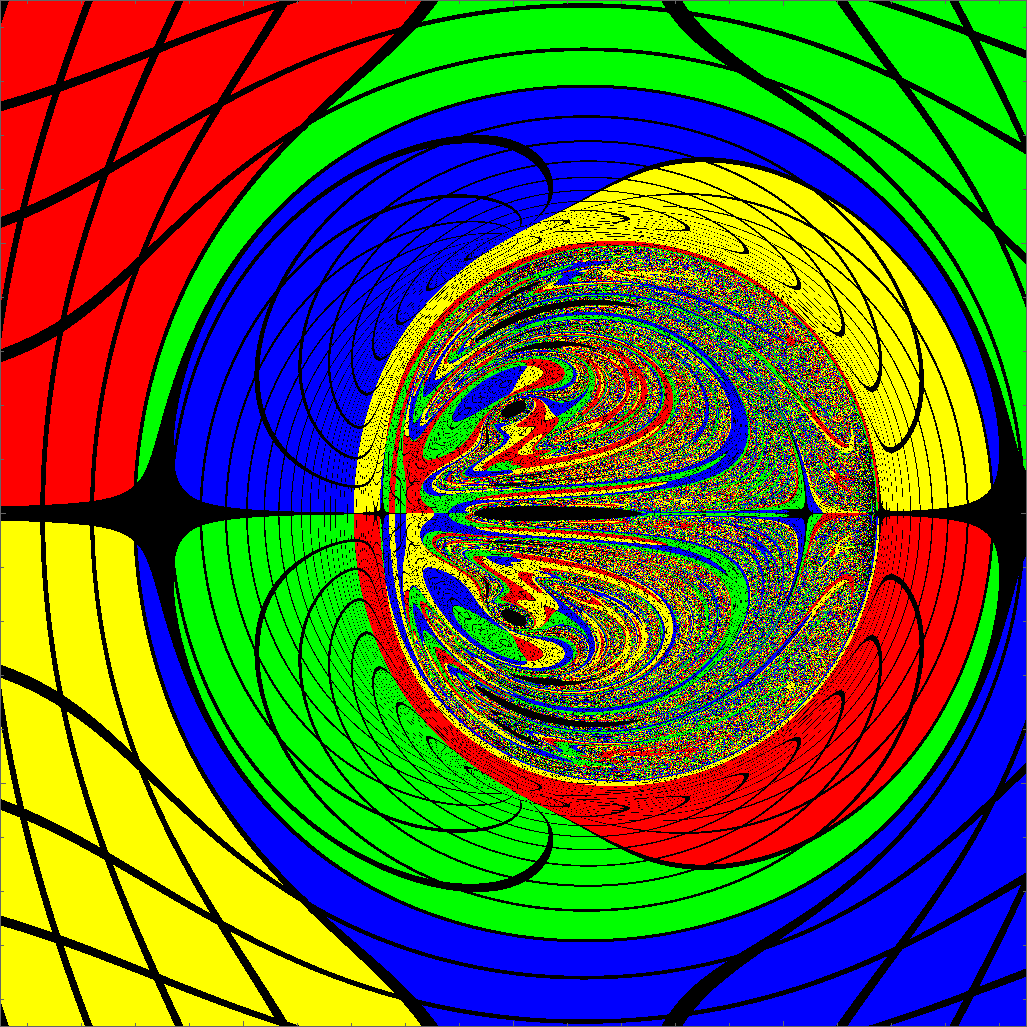}    
    \includegraphics[width=0.33\textwidth]{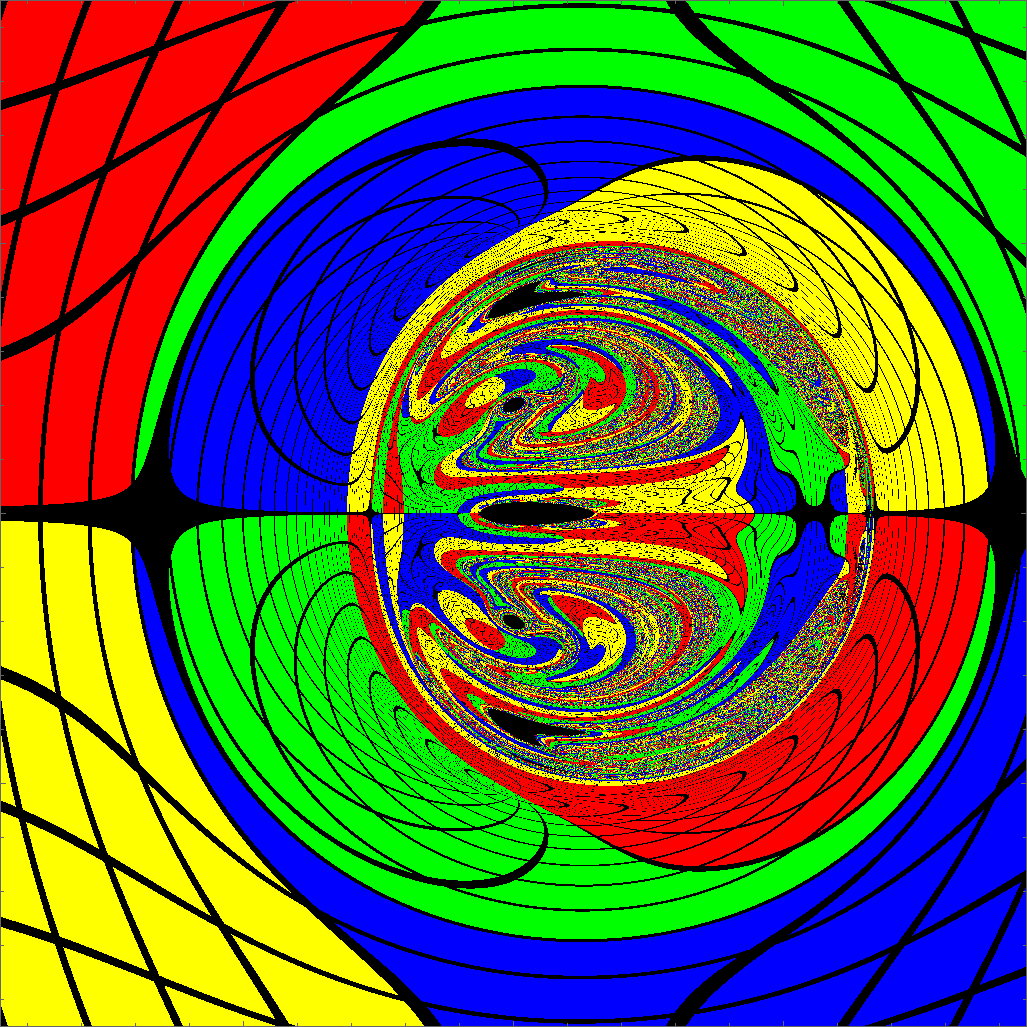}    
    \includegraphics[width=0.33\textwidth]{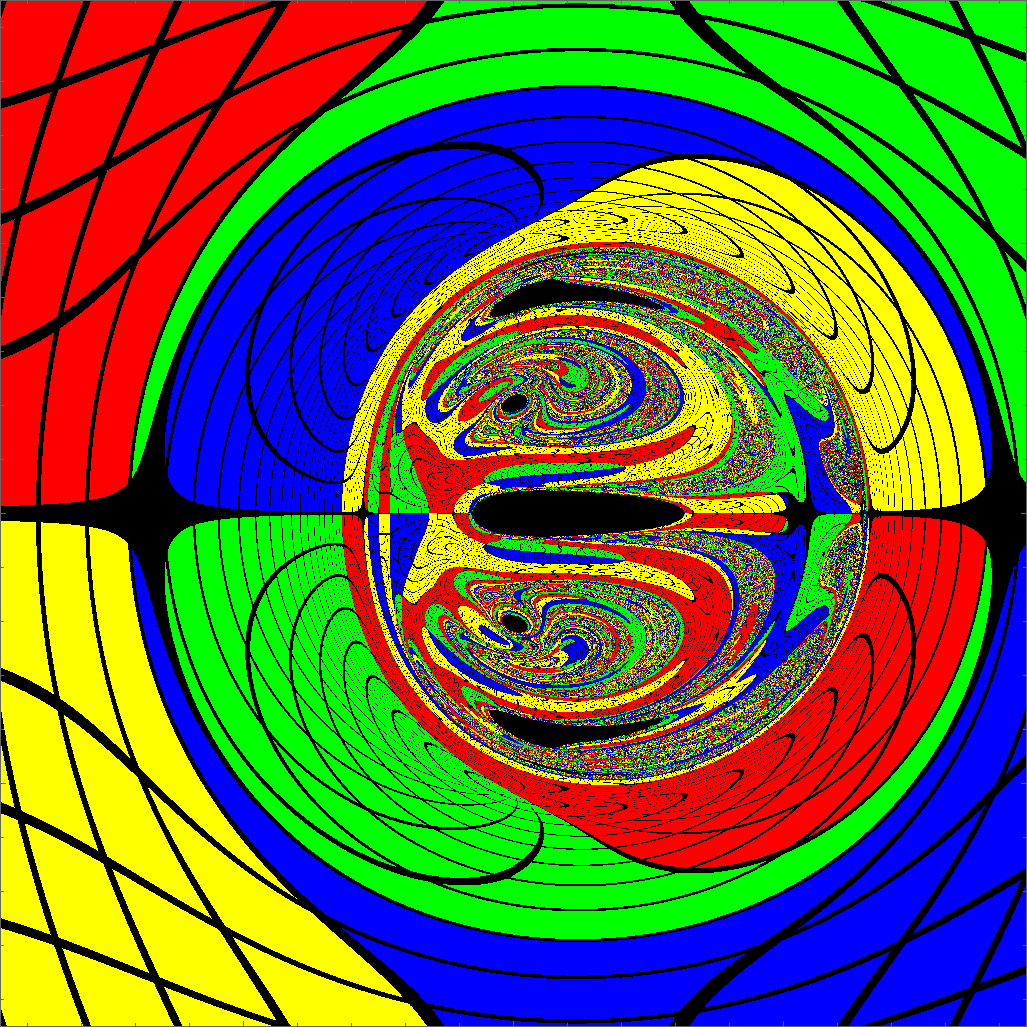}
    \caption{\small Examples of shadows which illustrate the transition between shadows for different Gaussian curvatures $\kappa\in\{-5,0,5\}$ and black hole horizon $r_H=0.01$. The left panel corresponds to configuration \textbf{I}$^{\,-5}_{\,0.01}$ with $\kappa=-5$, $\omega_{s}/\mu=0.607387$, $M\mu=0.890489$ and $q=0.999866$, the center panel corresponds to configuration \textbf{I}$^{\,0}_{\,0.01}$ with $\kappa=0$, $\omega_{s}/\mu=0.679241$, $M\mu=0.881991$ and $q=0.999875$, while the right panel corresponds to configuration \textbf{I}$^{\,5}_{\,0.01}$ with $\kappa=5$, $\omega_{s}/\mu=0.731639$, $M\mu=0.883209$ and $q=0.999733$. To clarify the interpretation of the colour components in the images and their deformation, a color legend and a grid are provided in Fig. \ref{fig:ColorSph}. Detailed physical quantities of the solution are provided in Table \ref{tab_7} in Appendix A.}
	\label{fig:4}
\end{figure} 

In all three geometries, there exists an inner ergoregion extending from the horizon ($R_{H}=0$) to a distance $R_{ER}$, which slightly depends on the Gaussian curvature. A notable distinction is observed when the curvature of the target space is zero; the equatorial extension of the inner ergoregion is minimal, $R_{ER} \in [0, \, 0.003]$, in contrast to the extensions, $R_{ER} \in [0, \, 0.004]$ and $R_{ER} \in [0, \, 0.005]$, for negative and positive curvatures, respectively. No light ring is placed in any of these inner ergoregions.

All three configurations feature an ergotorus, corresponding to zero Gaussian curvature with the minor equatorial section, $R_{LR} \in [0.063, \, 0.462]$. In the other two cases, the ergotorus possesses a slightly larger width: $R_{ER} \in [0.045, \, 0.474]$ for negative curvature and $R_{ER} \in [0.043, \, 0.445]$ for positive curvature. Each ergotorus houses a stable light ring that co-rotates with the black hole, irrespective of the Gaussian curvature of the target space.

Regardless of the magnitude of the Gaussian curvature, each configuration has two inner unstable light rings located near each other, positioned between the two ergoregions. The innermost of these rings co-rotates with the black hole, while the second larger ring rotates in the opposite direction. A distinctive feature is observed in the ergotorus of the solution with zero Gaussian curvature, where the stable light ring has the largest radius, $R_{LR}\simeq0.196$, in contrast to the smaller-radius stable rings, $R_{LR}\simeq0.155$, for positive curvature, and $R_{LR}\simeq0.186$, for negative curvature. Outside the ergotorus is positioned the outermost fourth unstable light ring, which rotates in the opposite direction to the black hole.

The shadows of the considered solutions are depicted in Fig. \ref{fig:4}. Under negative Gaussian curvature, the shadow is densely filled with chaotic patterns that exhibit slight changes in both structure and location at zero and positive curvature. A significant feature at this nearly extreme value of the normalized charge in scalar hair black hole configurations is the emergence of multiple highly elongated shadows. Specifically, under negative Gaussian curvature, the shadow displays 11 distinguishable simply-connected miniature dark regions. This number decreases to 7 shadow regions at zero and positive curvature. During this transition, four dark regions merge into two slightly more extensive regions, while two other dark regions disappear. This change results in a noticeable, albeit small, reduction in chaotic patterns. Furthermore, the increase in curvature convincingly demonstrated an enlargement in the width and overall size of the multiple dark areas formed, contributing to the entire shadow image.

\subsubsection{Model II, $r_H=0.01$, $q\simeq0.997$: Shadows with large chaotic regions.}

Analyzing the influence of the normalized charge, $q$, on shadow formation, ergoregions, and the system of light rings, we move through the $M-\omega_{s}$ space, shown in Fig. \ref{fig:kappa}, to the right while maintaining a constant horizon radius, $r_{H}=0.01$. We select configurations \textbf{II}$^{\,\kappa}_{\,0.01}$, with $\kappa\in\{-5,0,5\}$, all with slightly smaller values of the normalized charge, roughly $q\simeq0.997$. The corresponding physical parameters for these configurations are provided in Table \ref{tab_7} in Appendix A.

Each configuration with a different Gaussian curvature has one ergoregion and four light rings. Among these light rings, there is always one that remains stable, contributing to the observation of chaotic regions in shadow images. The peculiarity of these configurations is their ability to form multiple non-simply connected images, which shows significant chaotic behaviour in trajectories of light orbits near the black hole horizon. 

The ergoregion of the \textbf{II$^{\,-5}_{\,0.01}$} configuration has the most spacious equatorial cross-section, extending from the black hole's event horizon ($R_{H}=0$) to $R_{ER}\simeq0.410$. The outermost light ring at $R_{LR}\simeq0.674$, is unstable and rotates opposite to the black hole. Moving towards the event horizon, we pass through the ergosurface ($g_{tt}=0$) and discover a system of three equatorial light rings relatively close to the horizon. The outermost of these rings, at $R_{LR}\simeq0.071$, is stable and rotates in the direction of the black hole's rotation. The next inner ring, at $R_{LR}\simeq0.014$, is unstable and rotates oppositely to the black hole's rotation. The innermost light ring, closest to the horizon, is also unstable, rotates in the direction of the black hole's rotation, and has a radius corresponding to coordinate $R_{LR}\simeq0.009$. The positions of the light rings and the equatorial domain of ergoregion existence, expressed in terms of the dimensionless parameter $R$, are summarized in the following table.

\begin{center}
\setlength{\tabcolsep}{5.3pt} 
\renewcommand{\arraystretch}{1} 
\begin{tabular}{c*{10}{c}cc}
\hline
\hline
Configuration & Fig. & Ergoregions & $R_{ER}$ & LR & $R_{LR}$ & stability & $\eta$ & $g_{tt}$ & $d\varphi/dt$ & Chaos\\
\hline
\multirow{4}{*}{\textbf{II}$^{\,-5}_{\,0.01}$} & \multirow{4}{*}{\ref{fig:3}} & \multirow{4}{*}{1 ER} & \multirow{4}{*}{$[0, \; 0.410]$} & $h_-$ & 0.009 & unstable & $+$ & $+$ & $+$ & \multirow{4}{*}{Yes} \\
                      &                              &                       &                           & $h_+$ & 0.014 & unstable & $-$ & $+$ & $-$  \\
                      &                              &                       &                           & $h_+$ & 0.071 & stable & $+$ & $+$ & $+$  \\
                      &                              &                       &                           & $h_+$ & 0.674 & unstable & $-$ & $-$ & $-$  \\
\hline
\multirow{4}{*}{\textbf{II}$^{\,0}_{\,0.01}$} & \multirow{4}{*}{\ref{fig:3}} & \multirow{4}{*}{1 ER} & \multirow{4}{*}{$[0, \; 0.387]$} & $h_-$ & 0.008 & unstable & $+$ & $+$ & $+$  & \multirow{4}{*}{Yes} \\
                      &                              &                       &                           & $h_+$ & 0.016 & unstable & $-$ & $+$ & $-$  \\
                      &                              &                       &                           & $h_+$ & 0.069 & stable & $+$ & $+$ & $+$  \\
                      &                              &                       &                           & $h_+$ & 0.657 & unstable & $-$ & $-$ & $-$  \\
\hline
\multirow{4}{*}{\textbf{II}$^{\,5}_{\,0.01}$} & \multirow{4}{*}{\ref{fig:3}} & \multirow{4}{*}{1 ER} & \multirow{4}{*}{$[0, \; 0.397]$} & $h_-$ & 0.007 & unstable & $+$ & $+$ & $+$  & \multirow{4}{*}{Yes} \\
                      &                              &                       &                           & $h_+$ & 0.020 & unstable & $-$ & $+$ & $-$  \\
                      &                              &                       &                           & $h_+$ & 0.063 & stable & $+$ & $+$ & $+$  \\
                      &                              &                       &                           & $h_+$ & 0.680 & unstable & $-$ & $-$ & $-$  \\
\hline\hline		
\end{tabular}
\end{center}

\begin{figure}
    \includegraphics[width=0.33\textwidth]{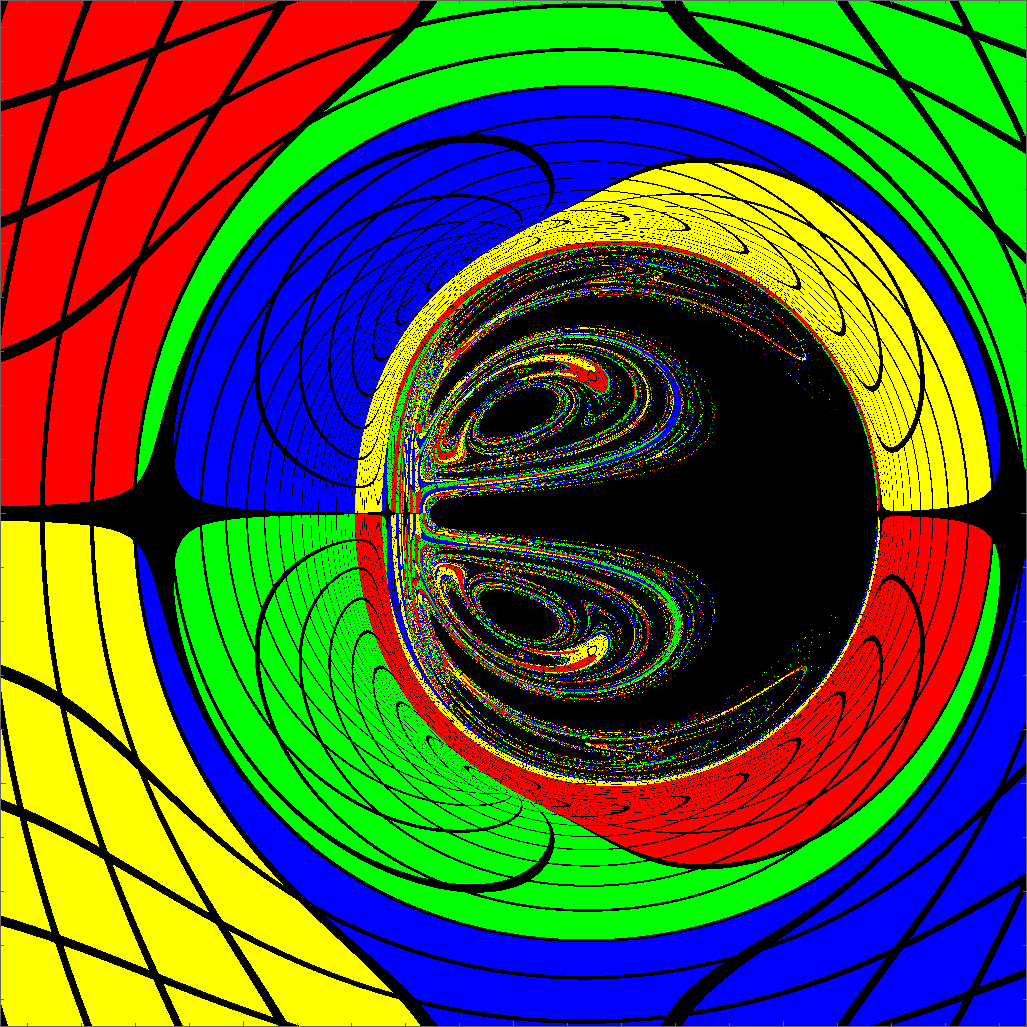}
    \includegraphics[width=0.33\textwidth]{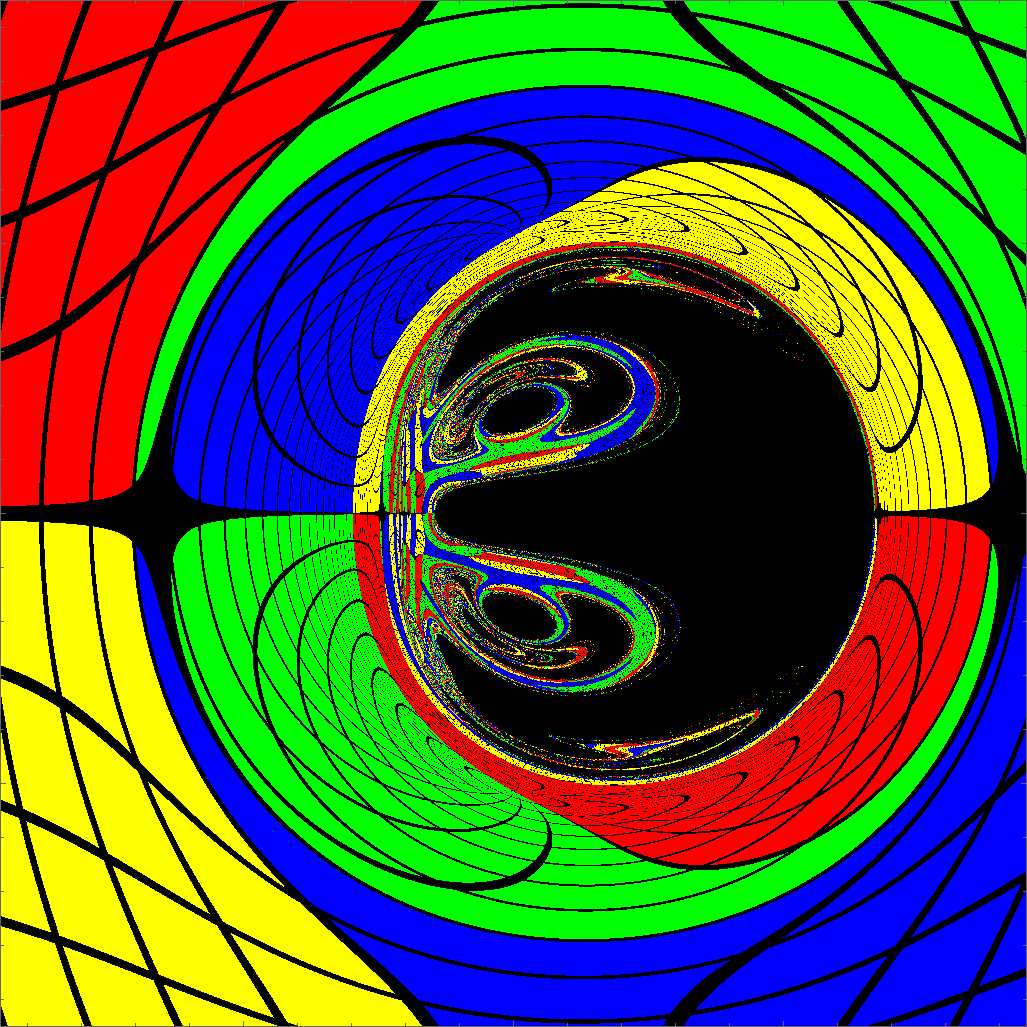}
    \includegraphics[width=0.33\textwidth]{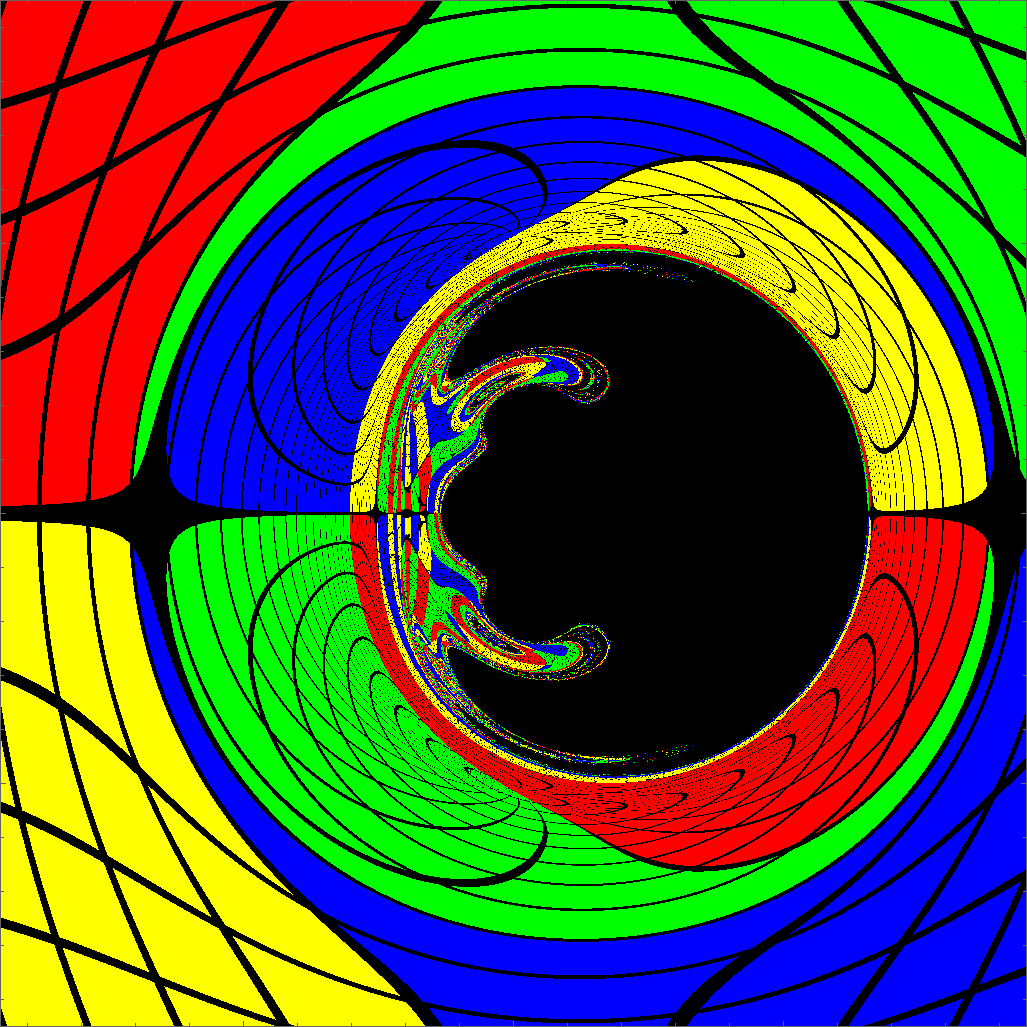}
    \caption{\small Examples of shadows which illustrate the transition between shadows for different Gaussian curvatures $\kappa\in\{-5,0,5\}$ and black hole horizon $r_H=0.01$. The left panel corresponds to configuration \textbf{II$^{\,-5}_{\,0.01}$} with $\kappa=-5$, $\omega_{s}/\mu=0.739809$, $M\mu=0.690217$ and $q=0.997229$, the center panel corresponds to configuration \textbf{II$^{\,0}_{\,0.01}$} with $\kappa=0$, $\omega_{s}/\mu=0.835272$, $M\mu=0.648229$ and $q=0.997293$, while the right panel corresponds to configuration \textbf{II$^{\,5}_{\,0.01}$} with $\kappa=5$, $\omega_{s}/\mu=0.821927$, $M\mu=0.742514$ and $q=0.997069$. To clarify the interpretation of the colour components in the images and their deformation, a color legend and a grid are provided in Fig. \ref{fig:ColorSph}. Detailed physical quantities of the solution are provided in Table \ref{tab_7} in Appendix A.}
	\label{fig:3}
\end{figure} 

Referring to Fig. \ref{fig:3}, we observe that an increase in the Gaussian curvature results in a noticeable expansion of the visible area of black hole shadows, maintaining the same normalized charge. For the most substantial negative value of the Gaussian curvature, $k=-5$, multiple nested ovals and non-simply connected images are observed, indicating a significant chaotic behavior in the trajectories of light orbits. As the Gaussian curvature increases to $k=0$, chaotic patterns diminish in size, forming distinguishable multiple-shadow images of the black hole. In the case of the minor event horizon, $r_{H}=0.01$, one reason for the reduction in chaotic patterns within the shadow images is the shrinking of the radius of stable light rings with increasing Gaussian curvature. Simultaneously, in the considered configurations, the outermost unstable light rings move away from the event horizon, and their radius further expands with the increasing curvature of the target space. This phenomenon is an essential condition for increasing the capture cross-section of the black holes, ultimately resulting in the formation of a larger shadow.

\subsubsection{Model III, $r_H=0.05$, $q\simeq0.994$: Larger size black hole shadows harbouring chaotic regions.}

Increasing the size of the event horizon to $r_{H}=0.05$, we consider configurations \textbf{III}$^{\,\kappa}_{\,0.05}$, with $\kappa\in\{-5,0,5\}$, highlighted in Fig. \ref{fig:kappa}. In this scenario, the selected configurations feature a normalized charge of $q\simeq0.994$. The physical quantities of these solutions are detailed in Table \ref{tab_7} in Appendix A, showcasing variations for different Gaussian curvatures of the target space. As in configurations \textbf{I} and \textbf{VII}, in configuration \textbf{III}, we observe the appearance of the same structure of four light rings and two ergoregions. The table below presents the technical characteristics of the geometric structures under consideration. 

\begin{center}
\setlength{\tabcolsep}{4.5pt} 
\renewcommand{\arraystretch}{1.1} 
\begin{tabular}{c*{10}{c}cc}
\hline
\hline
Configuration & Fig. & Ergoregions & $R_{ER}$ & LR & $R_{LR}$ & stability & $\eta$ & $g_{tt}$ & $d\varphi/dt$ & Chaos\\
\hline
\multirow{4}{*}{\textbf{III}$^{\,-5}_{\,0.05}$} & \multirow{4}{*}{\ref{fig:5}} &                          &                           & $h_-$ & 0.040 & unstable & $+$ & $+$ & $+$  & \multirow{4}{*}{Yes} \\
                        &                              & 1 ER & $[0, \; 0.057]$     & $h_+$ & 0.073 & unstable & $-$ & $-$ & $-$ & \\
                        &                              & 2 ER & $[0.095, \; 0.476]$ & $h_+$ & 0.274 & stable & $+$ & $+$ & $+$ & \\
                        &                              &                          &                          & $h_+$ & 0.727 & unstable & $-$ & $-$ & $-$ & \\
\hline
\multirow{4}{*}{\textbf{III}$^{\,0}_{\,0.05}$} & \multirow{4}{*}{\ref{fig:5}} &                          &                           & $h_-$ &  0.037 & unstable & $+$ & $+$ & $+$ & \multirow{4}{*}{Yes} \\
                        &                              & 1 ER & $[0, \; 0.047]$     & $h_+$ & 0.080 & unstable & $-$ & $-$ & $-$ & \\
                        &                              & 2 ER & $[0.157, \; 0.471]$ & $h_+$ & 0.297 & stable & $+$ & $+$ & $+$ & \\
                        &                              &                          &                          & $h_+$ & 0.739 & unstable & $-$ & $-$ & $-$ & \\
\hline
\multirow{4}{*}{\textbf{III}$^{\,5}_{\,0.05}$} & \multirow{4}{*}{\ref{fig:5}} &                          &                           & $h_-$ & 0.036 & unstable & $+$ & $+$ & $+$ & \multirow{4}{*}{Yes} \\
                        &                              & 1 ER & $[0, \; 0.045]$     & $h_+$ & 0.087 & unstable & $-$ & $-$ & $-$ & \\
                        &                              & 2 ER & $[0.238, \; 0.444]$ & $h_+$ & 0.333 & stable & $+$ & $+$ & $+$ & \\
                        &                              &                          &                          & $h_+$ & 0.757 & unstable & $-$ & $-$ & $-$ & \\
\hline\hline		
\end{tabular}
\end{center}

In configuration \textbf{III}, similar to configurations \textbf{I} and \textbf{VII}, we note the presence of the same structure, featuring four light rings and two ergoregions. However, in contrast to configurations \textbf{I} and \textbf{VII}, where both four light rings and two ergoregions coexist, configuration \textbf{III} exhibits a different arrangement -- the innermost unstable ring is located within the inner ergoregion. Transitioning into the zone between the two ergoregions, a single unstable light ring is found. Similar to solutions \textbf{I} and \textbf{VII}, the ergoregion also hosts a stable light ring. Extending beyond the ergoregion, we establish the existence of the fourth outermost unstable light ring.

Analyzing the dynamics of the light rings, we observe that, moving from the innermost to the outermost ring, the first co-rotates with the black hole, while subsequent rings alternate in spin direction. As a result, the fourth and final light ring counter-rotates with the black hole. When increasing the Gaussian curvature of the target space, we note a corresponding growth in the radius of light rings of the same type. Consequently, the system of light rings for positive Gaussian curvature exhibits the most extensive spatial distribution. At the same time, the equatorial region of both inner ergoregions and ergotoruses decreases with an increase in the Gaussian curvature across different configurations. 

\begin{figure}
    \includegraphics[width=0.33\textwidth]{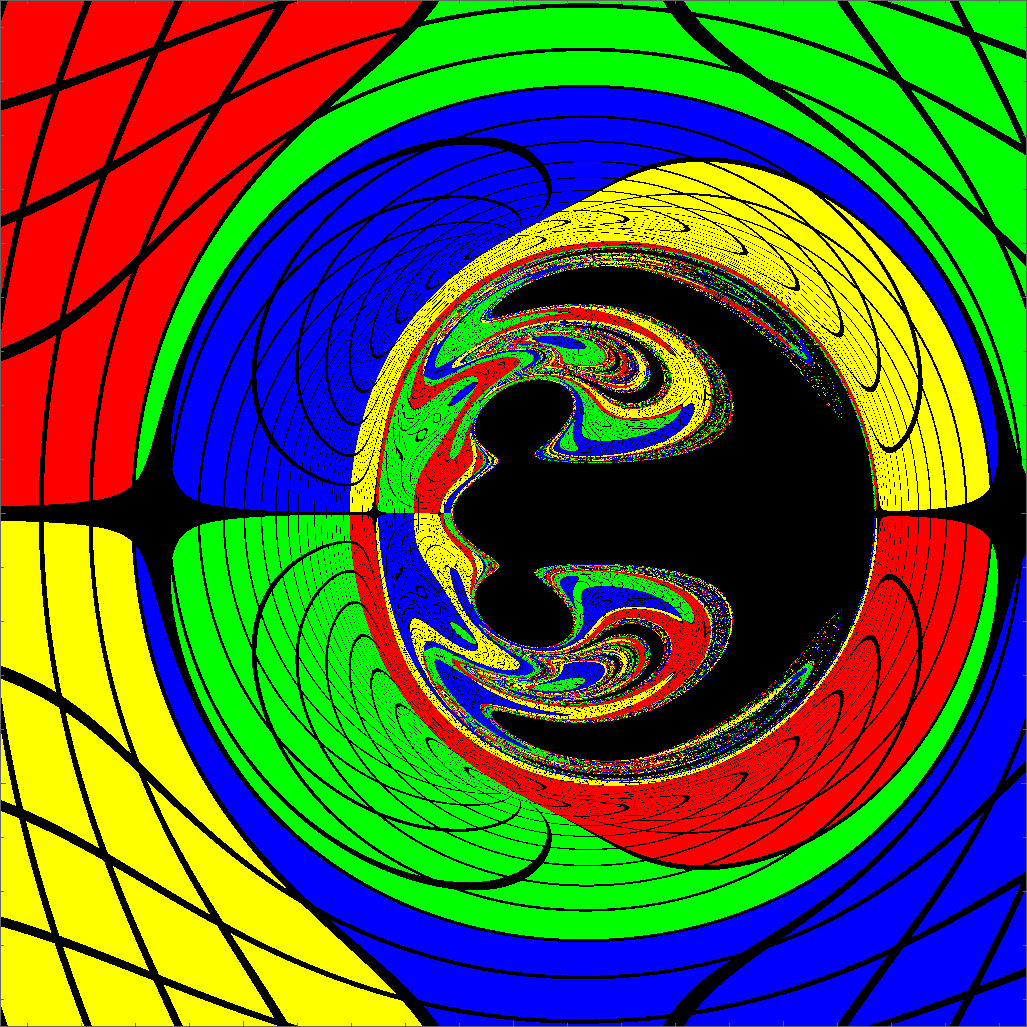}
    \includegraphics[width=0.33\textwidth]{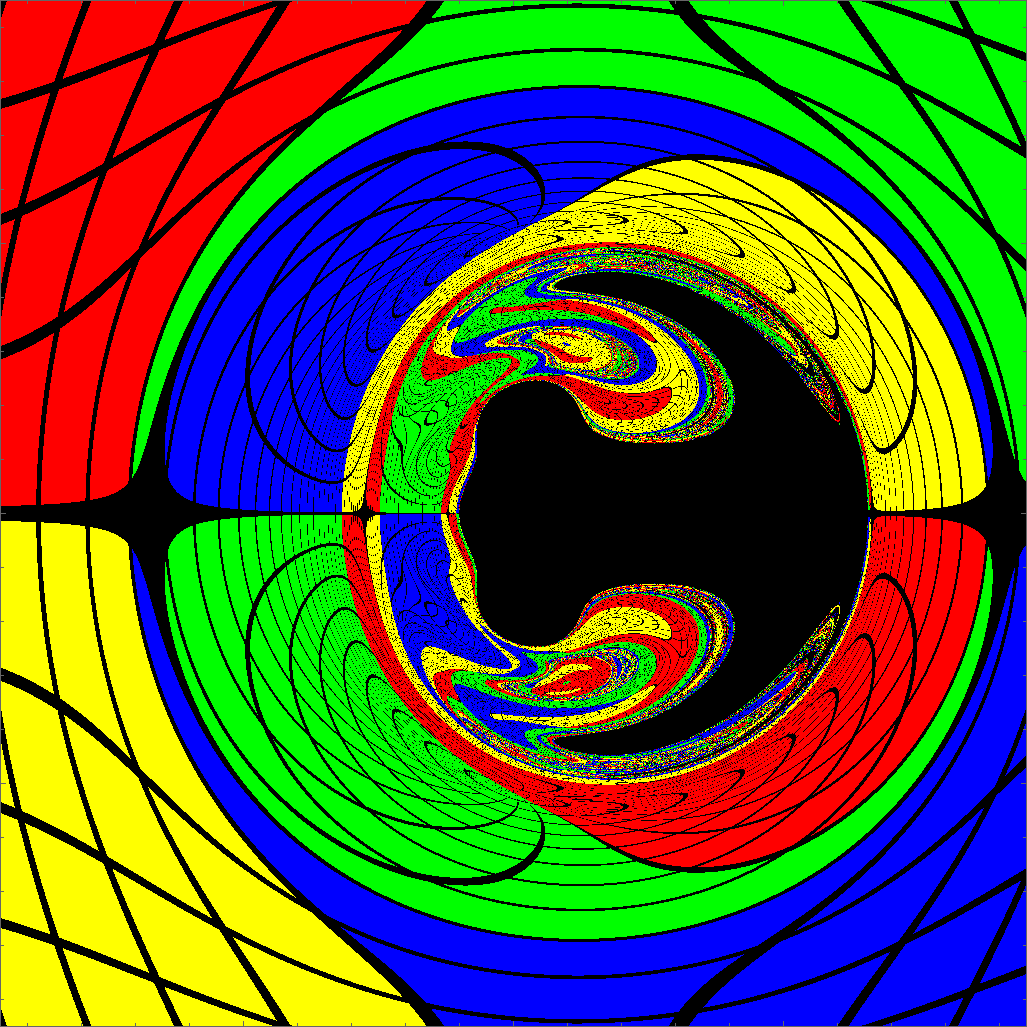}
    \includegraphics[width=0.33\textwidth]{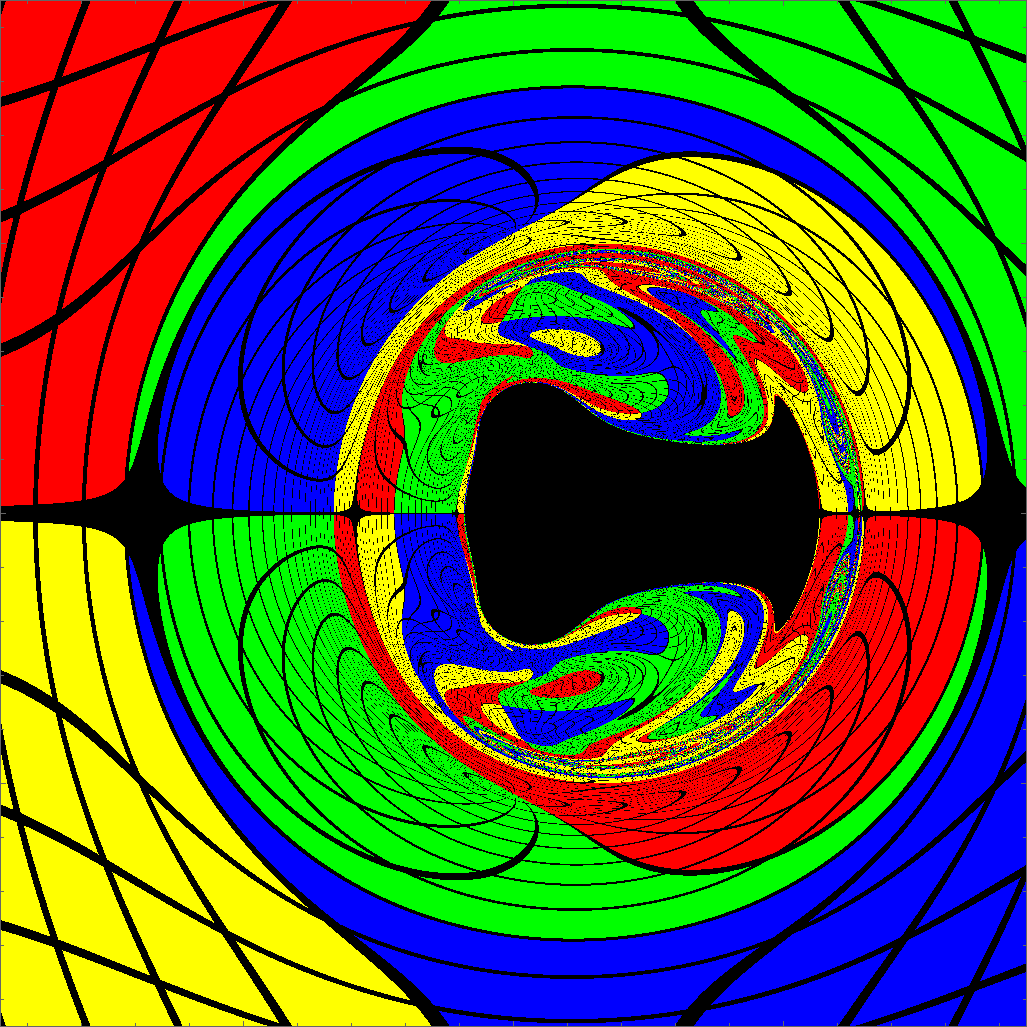}
    \caption{\small Examples of shadows which illustrate the transition between shadows for different Gaussian curvatures $\kappa\in\{-5,0,5\}$ and black hole horizon $r_H=0.05$. The left panel corresponds to configuration \textbf{III}$^{\,-5}_{\,0.05}$ with $\kappa=-5$, $\omega_{s}/\mu=0.653743$, $M\mu=0.886503$ and $q=0.994449$, the center panel corresponds to configuration \textbf{III}$^{\,0}_{\,0.05}$ with $\kappa=0$, $\omega_{s}/\mu=0.706437$, $M\mu=0.908153$ and $q=0.994490$, while the right panel corresponds to configuration \textbf{III}$^{\,5}_{\,0.05}$ with $\kappa=5$, $\omega_{s}/\mu=0.702254$, $M\mu=1.100262$ and $q=0.994709$. To clarify the interpretation of the colour components in the images and their deformation, a color legend and a grid are provided in Fig. \ref{fig:ColorSph}. Detailed physical quantities of the solution are provided in Table \ref{tab_7} in Appendix A.}
	\label{fig:5}
\end{figure}

At these values of the horizon and the normalized charge, we observe a significant change in the shape and size of the shadow when transitioning from negative to positive Gaussian curvature. At these values of the horizon and the normalized charge, we observe a significant change in the shape and size of the shadow when transitioning from negative to positive Gaussian curvature. Black hole shadows for configurations \textbf{III}$^{\,\kappa}_{\,0.05}$, with $\kappa\in\{-5,0,5\}$ are exposed in Fig. \ref{fig:5}. In instances of negative curvature, configuration \textbf{III}$^{\,-5}_{\,0.05}$ not only features a series of simply-connected dark regions but simultaneously exhibits chaotic patterns. At zero curvature of the target space, the individual simply-connected regions merge into a larger shadow that continues to show signs of randomness. In contrast to these cases, with positive curvature, the image of the shadow is fully connected, relatively more compact, and exhibits chaoticity of the scattered photon orbits outside the dark region of the shadow.

\subsubsection{Model IV, $r_H=0.1$, $q\simeq0.96$: Black hole shadows with decreasing chaotic regions.}

Let's consider the configurations \textbf{IV}$^{\,\kappa}_{\,0.1}$, with $\kappa\in\{-5,0,5\}$, highlighted in Fig. \ref{fig:kappa}, with an event horizon set at $r_{H}=0.1$. In these instances, the normalized charge of the selected solutions varies around $q\simeq0.96$. A notable feature of these configurations is that, for every value of the Gaussian curvature, a system of four light rings and one ergoregion exists. The positions of the light rings, the equatorial domain of the existence of the ergoregions, and additional information about their characteristics are presented in the following table. 

The equatorial domain of ergoregion existence is defined from the event horizon ($R_{H}=0$) to a distance $R_{ER}$, which decreases significantly as the Gaussian curvature of the target space increases. Therefore, at negative $\kappa=-5$, the relatively large width of the ergoregion, $R_{ER}\in[0, \, 0.439]$, encloses three of the inner light rings with radii $R_{LR}=0.061$, $R_{LR}=0.199$, and $R_{LR}=0.331$. The innermost and middle of these rings are unstable, while the third is stable, and all rotate in the direction of the black hole's rotation. In contrast, the outermost fourth light ring is located outside the ergoregion at $R_{LR}=0.724$. Moreover, it is unstable and rotates contrary to the black hole. In the absence of Gaussian curvature, the equatorial domain of the ergoregion is bounded by $R_{ER}\in[0, \, 0.147]$, signifying a shift of the outer boundary towards the event horizon. Additionally, the ergoregion hosts only one unstable light ring at $R_{LR}=0.057$, rotating in the direction of the black hole. Consequently, the other three light rings, situated at $R_{LR}=0.238$, $R_{LR}=0.395$, and $R_{LR}=0.742$, respectively, remain external to the ergoregion, each exhibiting retrograde rotation concerning the black hole. This pattern persists even in the presence of positive Gaussian curvature. The ergoregion contracts, narrowing to $R_{ER}\in[0, \, 0.133]$, housing only one unstable light ring at $R_{LR}=0.055$, which rotates prograde to the black hole. All remaining light rings are located outside the ergoregion, with radii corresponding to $R_{LR}=0.251$, $R_{LR}=0.501$, and $R_{LR}=0.759$, respectively, and exhibit retrograde rotation about the black hole. Generally, the third light ring from the inside out is consistently stable for any Gaussian curvature. Furthermore, as the Gaussian curvature increases, the innermost ring moves closer to the black hole, while the other three move away.
\begin{center}
\setlength{\tabcolsep}{5.3pt} 
\renewcommand{\arraystretch}{1} 
\begin{tabular}{c*{10}{c}cc}
\hline
\hline
Configuration & Fig. & Ergoregions & $R_{ER}$ & LR & $R_{LR}$ & stability & $\eta$ & $g_{tt}$ & $d\varphi/dt$ & Chaos \\
\hline
\multirow{4}{*}{\textbf{IV}$^{\,-5}_{\,0.1}$} & \multirow{4}{*}{\ref{fig:6}} & \multirow{4}{*}{1 ER} & \multirow{4}{*}{$[0, \; 0.439]$} & $h_-$ & 0.061 & unstable & $+$ & $+$ & $+$ & \multirow{4}{*}{Yes} \\
                      &                              &                       &                           & $h_+$ & 0.199 & unstable & $+$ & $+$ & $+$ & \\
                      &                              &                       &                           & $h_+$ & 0.331 & stable & $+$ & $+$ & $+$ &  \\
                      &                              &                       &                           & $h_+$ & 0.724 & unstable & $-$ & $-$ & $-$ & \\
\hline
\multirow{4}{*}{\textbf{IV}$^{\,0}_{\,0.1}$} & \multirow{4}{*}{\ref{fig:6}} & \multirow{4}{*}{1 ER} & \multirow{4}{*}{$[0, \; 0.147]$} & $h_-$ & 0.057 & unstable & $+$ & $+$ & $+$ & \multirow{4}{*}{Yes}\\
                      &                              &                       &                           & $h_+$ & 0.238 & unstable & $-$ & $-$ & $-$ & \\
                      &                              &                       &                           & $h_+$ & 0.395 & stable & $-$ & $-$ & $-$ & \\
                      &                              &                       &                           & $h_+$ & 0.742 & unstable & $-$ & $-$ & $-$ & \\
\hline
\multirow{4}{*}{\textbf{IV}$^{\,5}_{\,0.1}$} & \multirow{4}{*}{\ref{fig:6}} & \multirow{4}{*}{1 ER} & \multirow{4}{*}{$[0, \; 0.133]$} & $h_-$ & 0.055 & unstable & $+$ & $+$ & $+$ & \multirow{4}{*}{Yes} \\
                      &                              &                       &                           & $h_+$ & 0.251 & unstable & $-$ & $-$ & $-$ & \\
                      &                              &                       &                           & $h_+$ & 0.501 & stable & $-$ & $-$ & $-$ & \\
                      &                              &                       &                           & $h_+$ & 0.759 & unstable & $-$ & $-$ & $-$ & \\
\hline\hline		
\end{tabular}
\end{center}
\begin{figure}
    \includegraphics[width=0.33\textwidth]{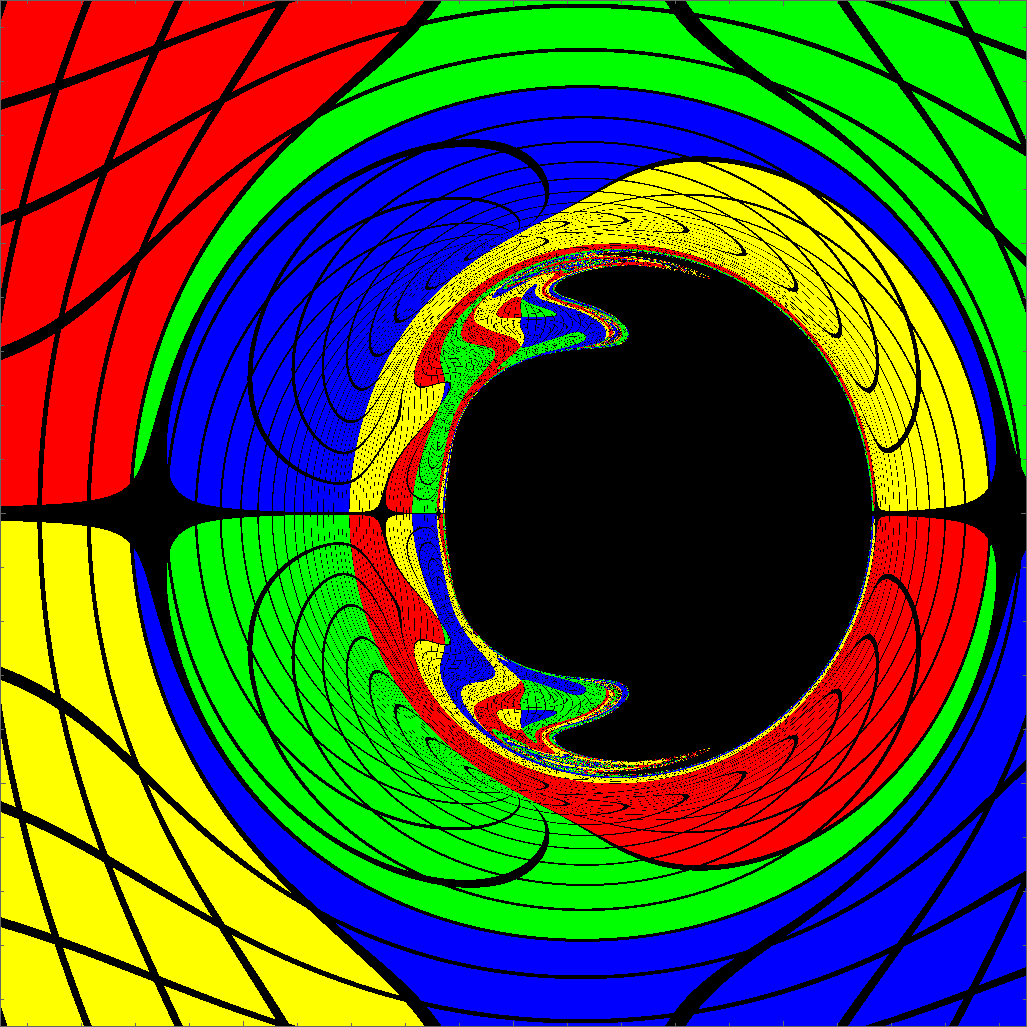}
    \includegraphics[width=0.33\textwidth]{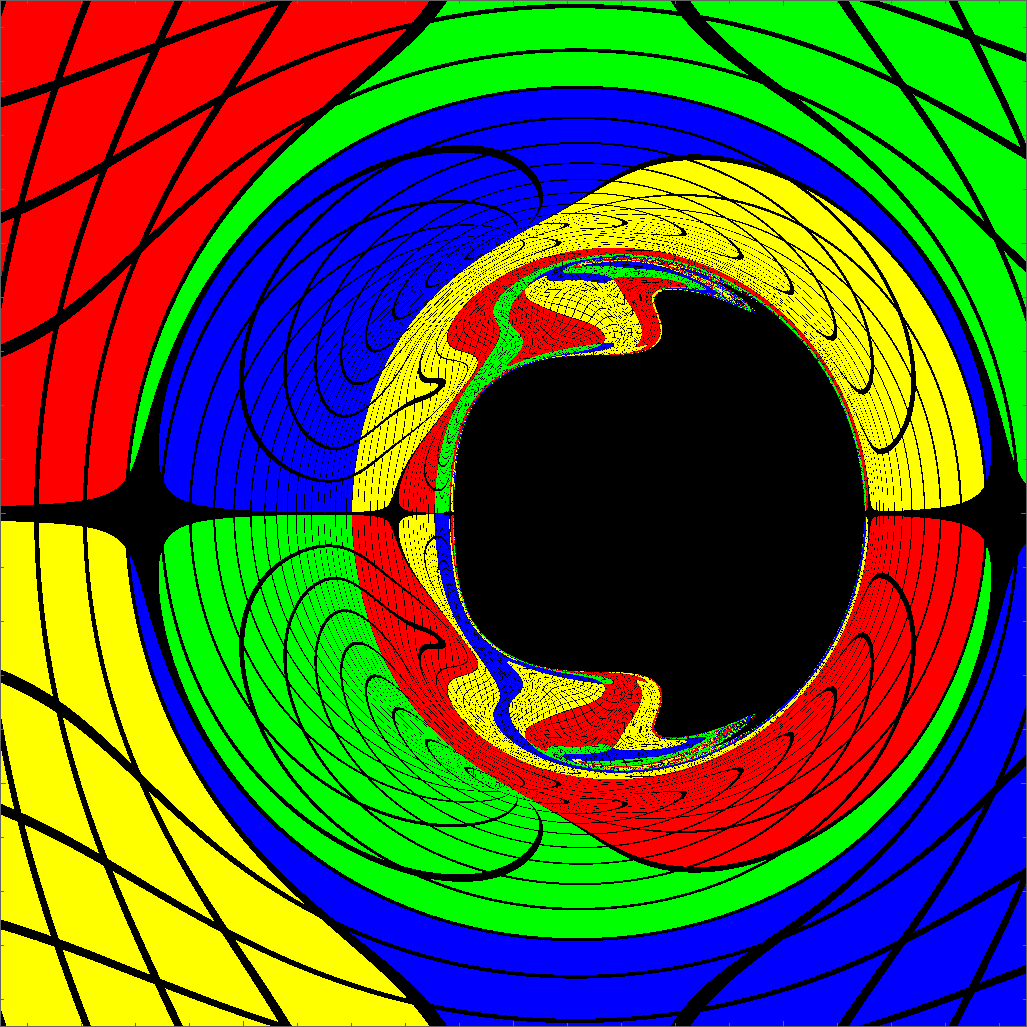}
    \includegraphics[width=0.33\textwidth]{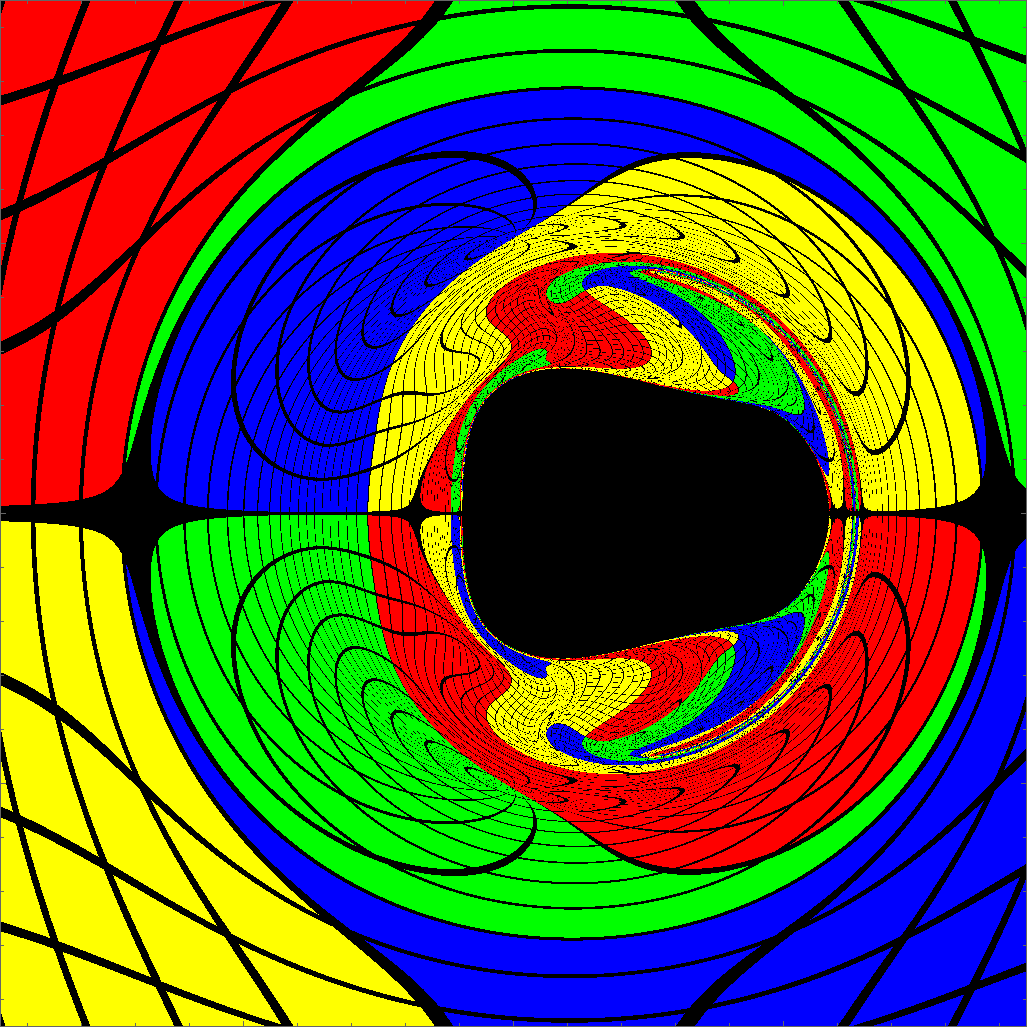}
    \caption{\small Examples of shadows which illustrate the transition between shadows for different Gaussian curvatures $\kappa\in\{-5,0,5\}$ and black hole horizon $r_H=0.1$. The left panel corresponds to configuration \textbf{IV}$^{\,-5}_{\,0.1}$ with $\kappa=-5$, $\omega_{s}/\mu=0.713790$, $M\mu=0.873847$ and $q=0.965279$, the center panel corresponds to configuration \textbf{IV}$^{\,0}_{\,0.1}$ with $\kappa=0$, $\omega_{s}/\mu=0.738499$, $M\mu=1.00043$ and $q=0.964477$, while the right panel corresponds to configuration \textbf{IV}$^{\,5}_{\,0.1}$ with $\kappa=5$, $\omega_{s}/\mu=0.748666$, $M\mu=1.16102$ and $q=0.968930$. To clarify the interpretation of the colour components in the images and their deformation, a color legend and a grid are provided in Fig. \ref{fig:ColorSph}. Detailed physical quantities of the solution are provided in Table \ref{tab_7} in Appendix A.}
	\label{fig:6}
\end{figure}

While stable light rings persist in the geometry of the considered configurations \textbf{IV}$^{\,\kappa}_{\,0.1}$, with $\kappa\in\{-5,0,5\}$, for the given value of the normalized charge, chaotic patterns in the shadow images (as shown in Fig. \ref{fig:6}) are notably lacking. One distinctive feature in the formed shadows is the emergence of tooth-like regions around the polar parts of the shadow contour. With an increase in Gaussian curvature, these regions gradually diminish, causing the images of the chaotically scattered orbits to shift from the inner part of the shadow towards its boundary. Accordingly, the chaotic patterns manifest at positive curvature in a thin, crescent-shaped region outside the shadow boundary. It is worth noting that the situation with configuration \textbf{III}$^{\,5}_{\,0.05}$ is analogous. This observation emphasizes that solutions with positive Gaussian curvatures result in shadows with considerably smaller areas and fewer chaotic patterns compared to those with zero or negative curvature of the target space.

\subsubsection{Model V, $r_H=0.2$, $q\simeq0.85$: Single ergoregion models with deformed quasi-circular shadows.}

\begin{figure}
    \includegraphics[width=0.33\textwidth]{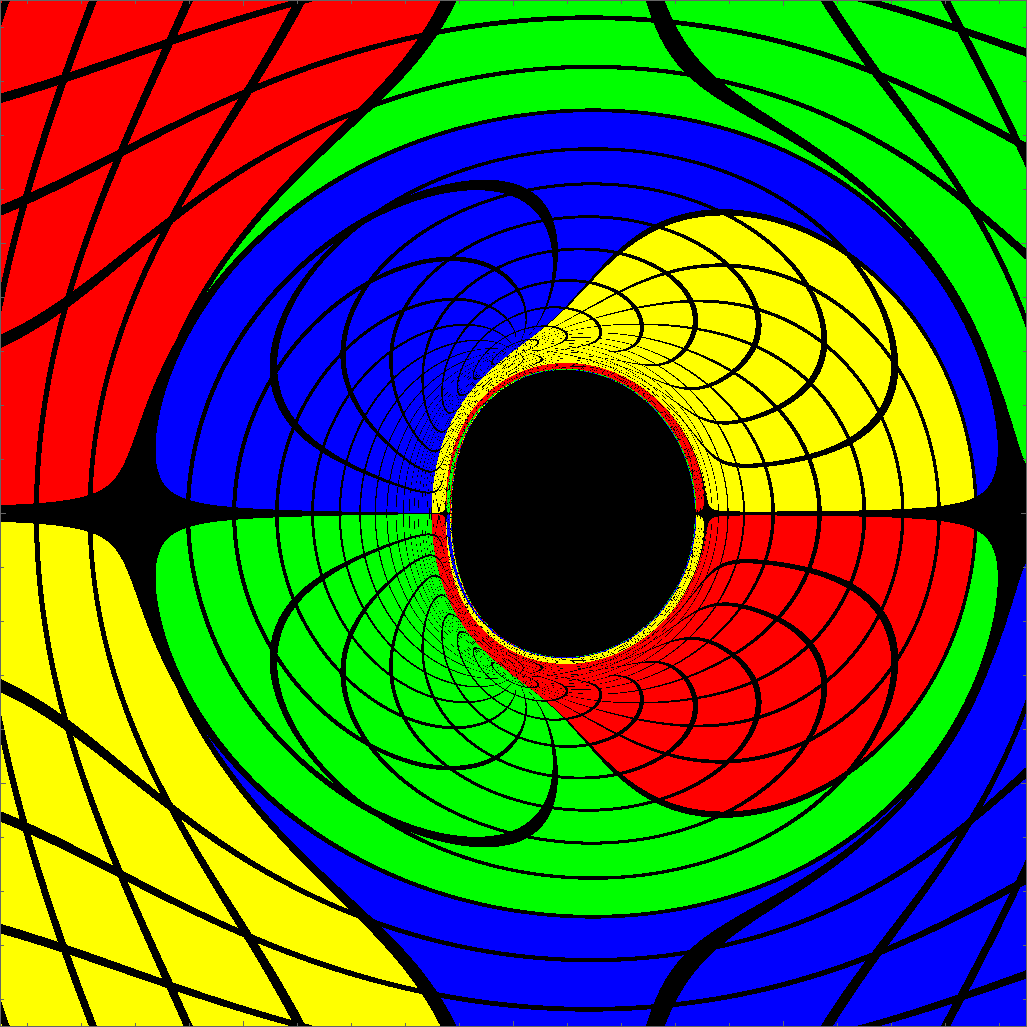}
    \includegraphics[width=0.33\textwidth]{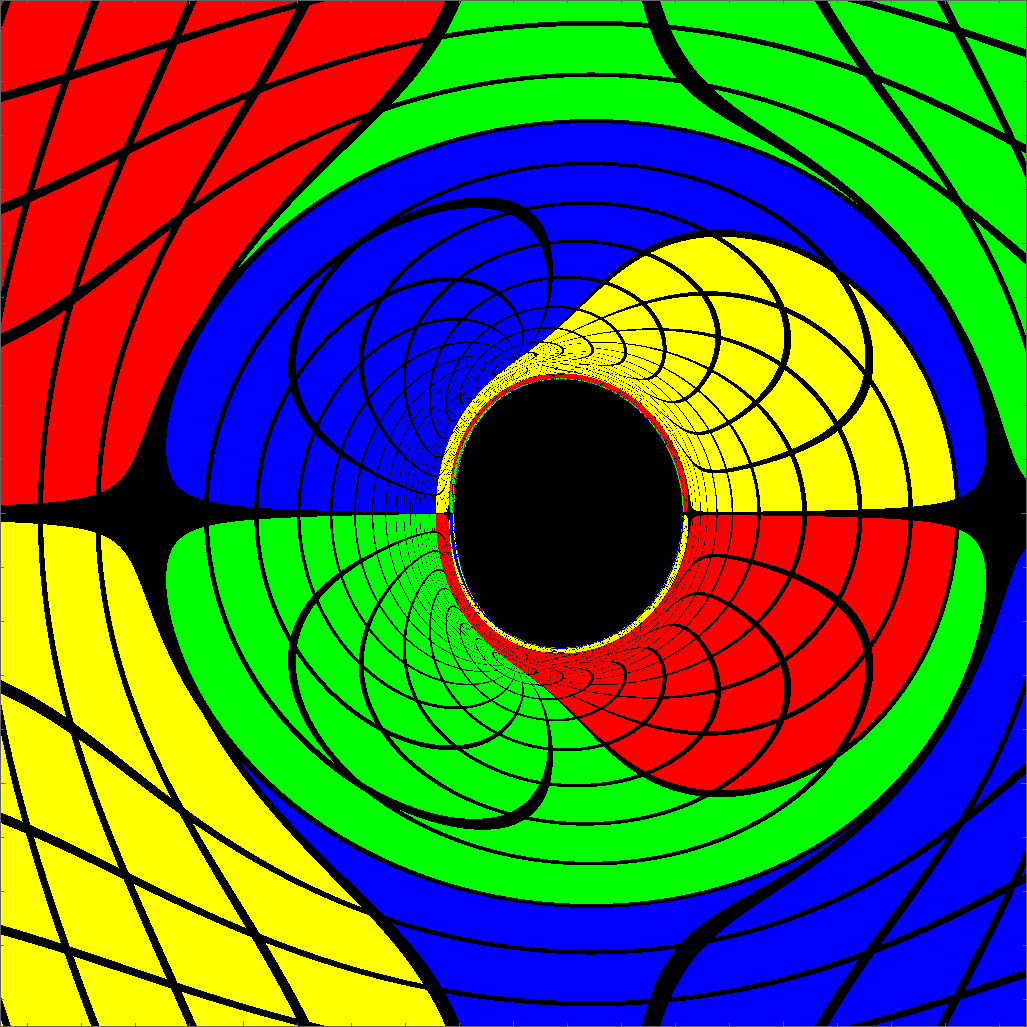}
    \includegraphics[width=0.33\textwidth]{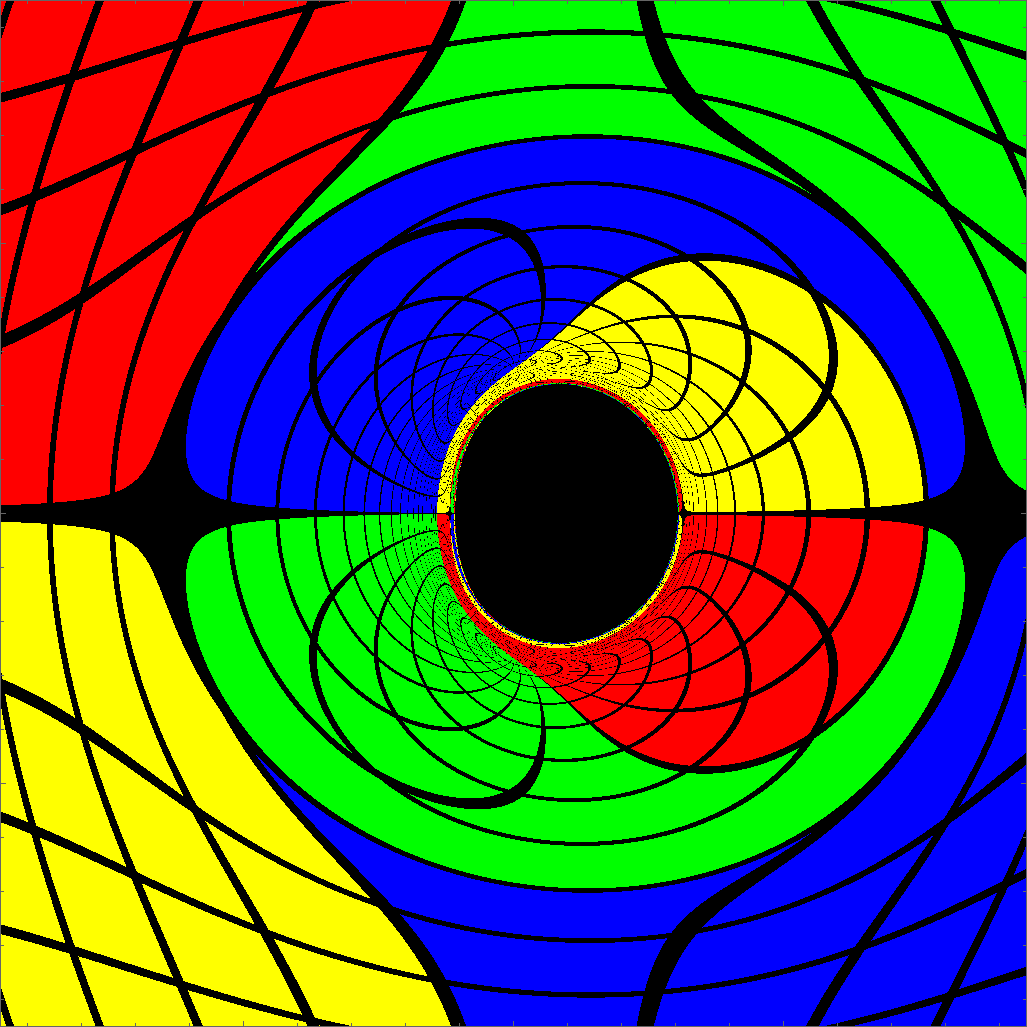}
    \caption{\small Examples of shadows which illustrate the transition between shadows for different Gaussian curvatures $\kappa\in\{-5,0,5\}$ and black hole horizon $r_H=0.2$. The left panel corresponds to configuration \textbf{V}$^{\,-5}_{\,0.2}$ with $\kappa=-5$, $\omega_{s}/\mu=0.883526$, $M\mu=0.847374$ and $q=0.848888$, the center panel corresponds to configuration \textbf{V}$^{\,0}_{\,0.2}$ with $\kappa=0$, $\omega_{s}/\mu=0.895538$, $M\mu=0.878726$ and $q=0.850778$, while the right panel corresponds to configuration \textbf{V}$^{\,5}_{\,0.2}$ with $\kappa=5$, $\omega_{s}/\mu=0.907762$, $M\mu=0.884665$ and $q=0.843196$. To clarify the interpretation of the colour components in the images and their deformation, a color legend and a grid are provided in Fig. \ref{fig:ColorSph}. Detailed physical quantities of the solution are provided in Table \ref{tab_7} in Appendix A.}
	\label{fig:7}
\end{figure}

Last but not least, the table below provides data on the characteristics of the ergoregions and the light ring system for configurations \textbf{V}$^{\,\kappa}_{\,0.2}$, where $\kappa\in\{-5,0,5\}$ (as indicated in Fig. \ref{fig:kappa}). These configurations correspond to an event horizon of $r_{H}=0.2$, and a normalized charge $q\simeq0.845$. Corresponding shadow images are depicted in Fig. \ref{fig:7}.

\begin{center}
\setlength{\tabcolsep}{5.3pt} 
\renewcommand{\arraystretch}{1} 
\begin{tabular}{c*{10}{c}cc}
\hline
\hline
Configuration & Fig. & Ergoregions & $R_{ER}$ & LR & $R_{LR}$ & stability & $\eta$ & $g_{tt}$ & $d\varphi/dt$ & Chaos \\
\hline
\multirow{2}{*}{\textbf{V}$^{\,-5}_{\,0.2}$} & \multirow{2}{*}{\ref{fig:7}} & \multirow{2}{*}{1 ER} & \multirow{2}{*}{$[0, \; 0.248]$} & $h_-$ & 0.095 & unstable & $+$ & $+$ & $+$ & \multirow{2}{*}{No} \\
                      &                              &                       &                           & $h_+$ & 0.471 & unstable & $-$ & $-$ & $-$ & \\
\hline
\multirow{2}{*}{\textbf{V}$^{\,0}_{\,0.2}$} & \multirow{2}{*}{\ref{fig:7}} & \multirow{2}{*}{1 ER} & \multirow{2}{*}{$[0, \; 0.252]$} & $h_-$ & 0.094 & unstable & $+$ & $+$ & $+$ & \multirow{2}{*}{No} \\
                      &                              &                       &                           & $h_+$ & 0.478 & unstable & $-$ & $-$ & $-$ & \\
\hline
\multirow{2}{*}{\textbf{V}$^{\,5}_{\,0.2}$} & \multirow{2}{*}{\ref{fig:7}} & \multirow{2}{*}{1 ER} & \multirow{2}{*}{$[0, \; 0.258]$} & $h_-$ & 0.093 & unstable & $+$ & $+$ & $+$ & \multirow{2}{*}{No} \\
                      &                              &                       &                           & $h_+$ & 0.488 & unstable & $-$ & $-$ & $-$ & \\
\hline\hline		
\end{tabular}
\end{center}

The considered configurations stand out from those previously studied in this paper due to the presence of a single ergoregion. Interestingly, these configurations exhibit a unique feature -- a system with two unstable light rings for all Gaussian curvatures of the target space. This distinctive characteristic draws comparisons with Kerr's rotating black hole, known for having one ergoregion and a system of two counter-rotating unstable light rings.

A notable feature of the examined configurations is that, as the Gaussian curvature increases, the equatorial domain of the ergoregions negligibly expands from $R_{ER}\in[0, \, 0.248]$ for negative $\kappa$, through $R_{ER}\in[0, \, 0.252]$ for zero $\kappa$, to $R_{ER}\in[0, \, 0.258]$ for positive $\kappa$. Each region hosts only one unstable light ring, prograde rotating regarding the black hole's rotation. Beyond the ergoregions, each configuration allows for a second, outer unstable light ring retrograde rotating concerning the black hole.

Interestingly, with an increase in Gaussian curvature, the inner rings move slightly closer to the horizon, acquiring radii $R_{LR}=0.095$, $R_{LR}=0.094$, and $R_{LR}=0.093$ for negative, zero, and positive $\kappa$, respectively. Simultaneously, the outer rings move away from the black hole as photons adopt orbits with radii $R_{LR}=0.471$, $R_{LR}=0.478$, and $R_{LR}=0.488$, correspondingly for negative, zero, and positive $\kappa$.

The distinctive behaviour of unstable prograde and retrograde light rings results in a shift in the equatorial portions of the shadows from west to east with increasing Gaussian curvature. Notably, this effect occurs when adjusting the curvature of the target space while keeping a constant the normalized charge $q$ for each solution. On the contrary, the phenomenon intensifies with an increase in the frequency $\omega_{s}/\mu$, or equivalently, with a rise in the angular velocity of the horizon as the scalar field is synchronized with the rotation of the black hole. Furthermore, as depicted in Fig. \ref{fig:7}, negative Gaussian curvature of the target space leads to forming a shadow with a larger area than the case without curvature. Conversely, with positive Gaussian curvature, the shadow area decreases. The reduction in shadow size appears as an outcome of the proportionally smaller fraction of the horizon mass to the ADM mass. Nevertheless, despite the varied Gaussian curvature, the shadows maintain a D-like shape, a distinctive feature resembling a rotating Kerr black hole as observed from the equatorial plane.

\subsubsection{Model VI, $r_H=0.3$, $q\simeq0.65$: Single ergoregion models with quasi-circular shadows.}

\begin{figure}
    \includegraphics[width=0.33\textwidth]{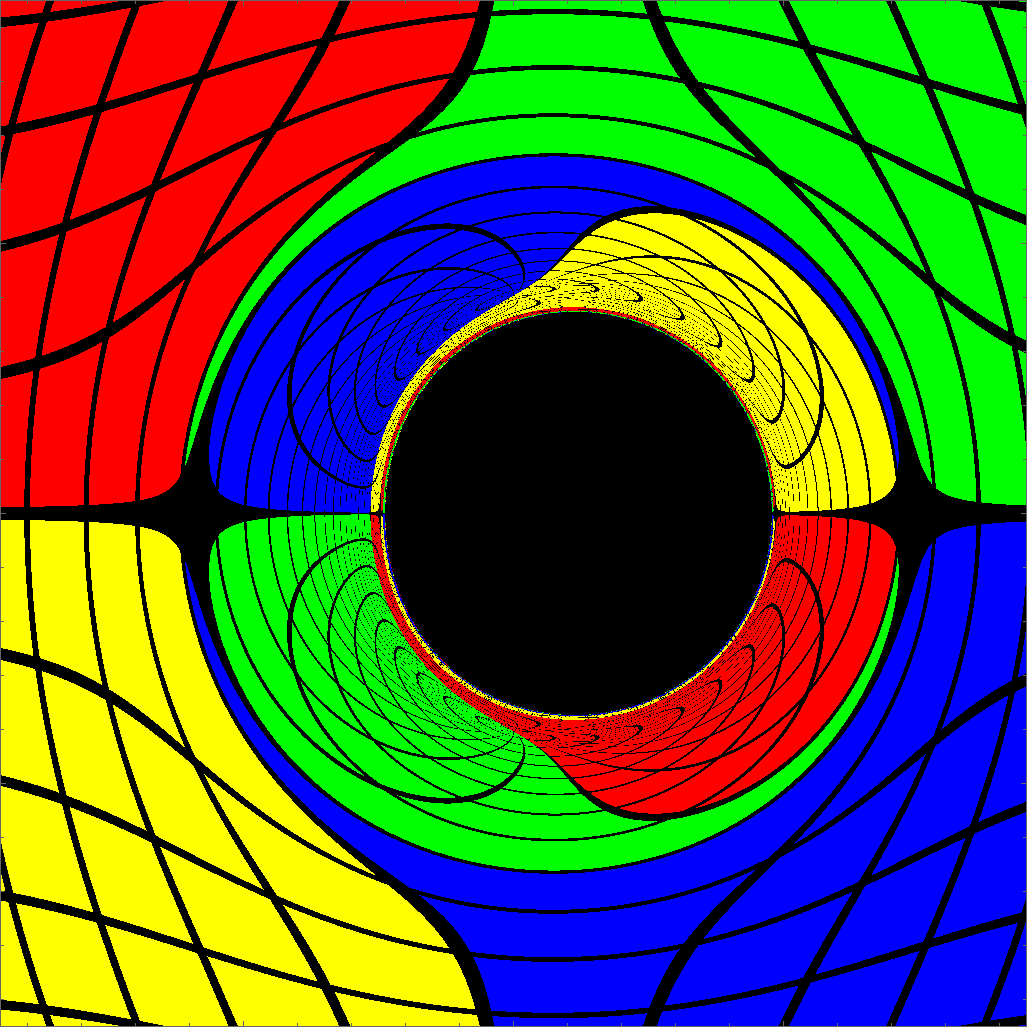}
    \includegraphics[width=0.33\textwidth]{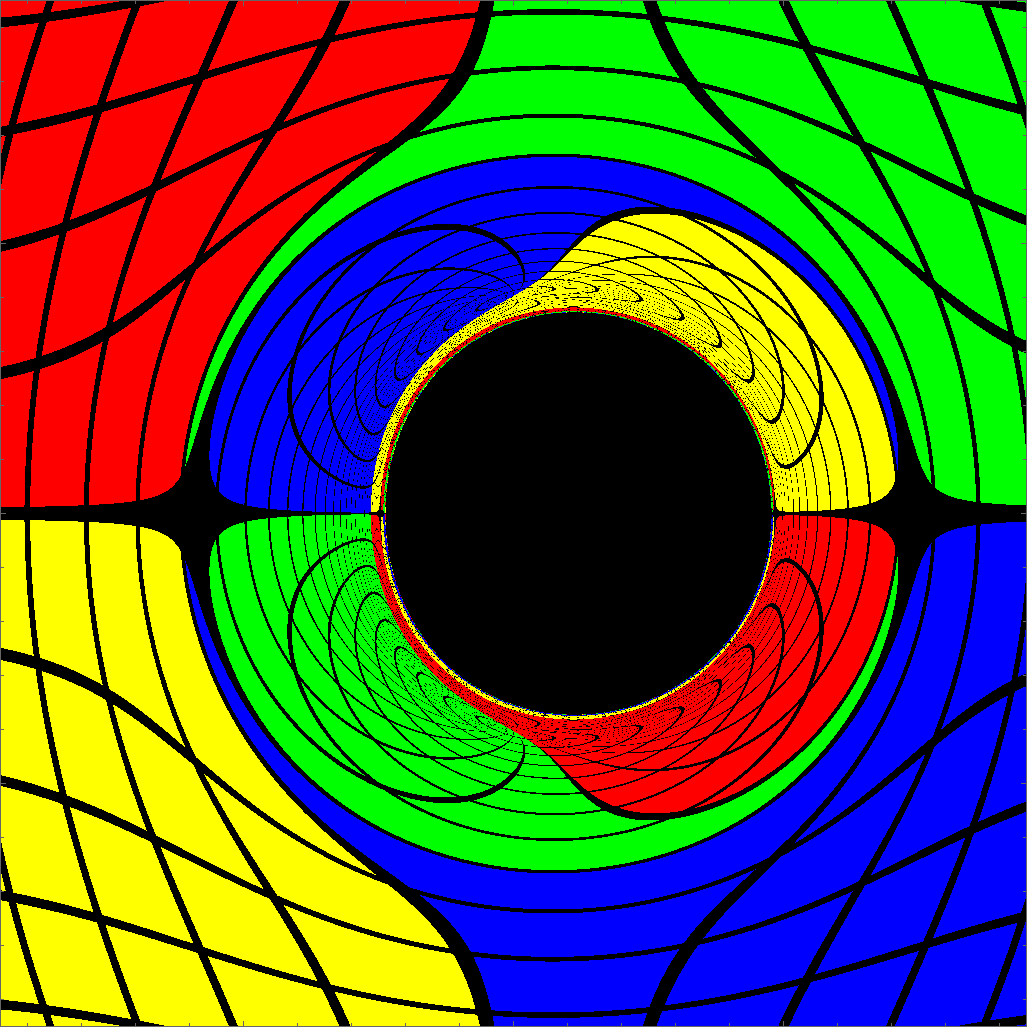}
    \includegraphics[width=0.33\textwidth]{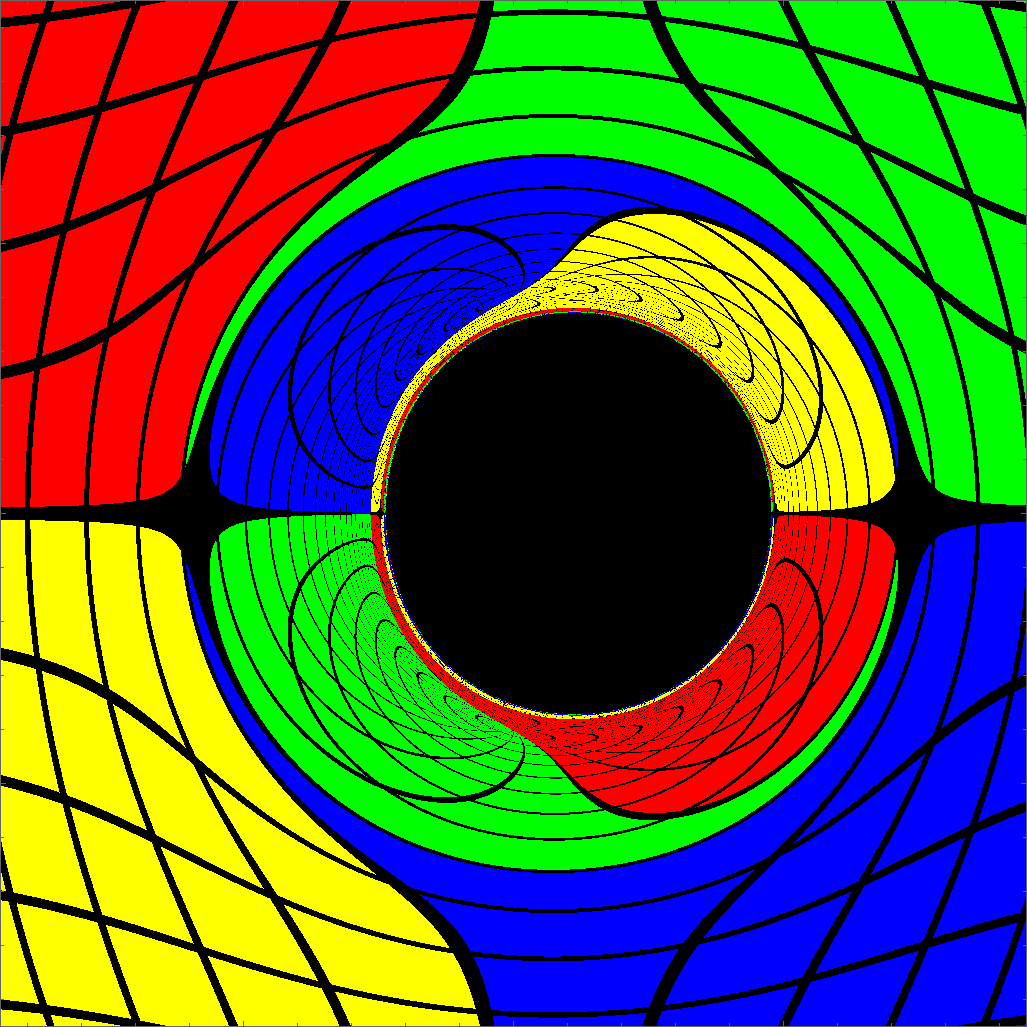}
    \caption{\small Examples of shadows which illustrate the transition between shadows for different Gaussian curvatures $\kappa\in\{-5,0,5\}$ and black hole horizon $r_H=0.3$. The left panel corresponds to configuration \textbf{VI}$^{\,-5}_{\,0.3}$ with $\kappa=-5$, $\omega_{s}/\mu=0.988000$, $M\mu=0.319008$ and $q=0.646883$, the center panel corresponds to configuration \textbf{VI}$^{\,0}_{\,0.3}$ with $\kappa=0$, $\omega_{s}/\mu=0.988000$, $M\mu=0.320010$ and $q=0.647791$, while the right panel corresponds to configuration \textbf{VI}$^{\,5}_{\,0.3}$ with $\kappa=5$, $\omega_{s}/\mu=0.988000$, $M\mu=0.321013$ and $q=0.648605$. The color setup and grid notation are explained in Fig. \ref{fig:ColorSph}, while detailed physical quantities are listed in Table \ref{tab_7} in Appendix A.}
	\label{fig:8}
\end{figure}

Finally, to study the structure of the system of light rings and ergoregions for an event horizon $r_{H}=0.3$, we choose configurations \textbf{VI}$^{\,\kappa}_{\,0.3}$, where $\kappa\in\{-5,0,5\}$ (as indicated in Fig. \ref{fig:kappa}). These configurations are characterized by a normalized charge $q\simeq0.647$. The corresponding characteristics of the light rings are presented in the table below, and shadow images are exposed in Fig. \ref{fig:8}.
\begin{center}
\setlength{\tabcolsep}{4.8pt} 
\renewcommand{\arraystretch}{1} 
\begin{tabular}{c*{10}{c}cc}
\hline
\hline
Configuration & Fig. & Ergoregions & $R_{ER}$ & LR & $R_{LR}$ & stability & $\eta$ & $g_{tt}$ & $d\varphi/dt$ & Chaos \\
\hline
\multirow{2}{*}{$-5.6$} & \multirow{2}{*}{\ref{fig:8}} & \multirow{2}{*}{1 ER} & \multirow{2}{*}{$[0, \; 0.1968]$} & $h_-$ & 0.1599 & unstable & $+$ & $+$ & $+$ & \multirow{2}{*}{No} \\
                      &                              &                       &                           & $h_+$ & 0.4306 & unstable & $-$ & $-$ & $-$ & \\
\hline
\multirow{2}{*}{\;\;\;$0.6$} & \multirow{2}{*}{\ref{fig:8}} & \multirow{2}{*}{1 ER} & \multirow{2}{*}{$[0, \; 0.1972]$} & $h_-$ & 0.1598 & unstable & $+$ & $+$ & $+$ & \multirow{2}{*}{No} \\
                      &                              &                       &                           & $h_+$ & 0.4310 & unstable & $-$ & $-$ & $-$ & \\
\hline
\multirow{2}{*}{\;\;\;$5.6$} & \multirow{2}{*}{\ref{fig:8}} & \multirow{2}{*}{1 ER} & \multirow{2}{*}{$[0, \; 0.1976]$} & $h_-$ & 0.1597 & unstable & $+$ & $+$ & $+$ & \multirow{2}{*}{No} \\
                      &                              &                       &                           & $h_+$ & 0.4314 & unstable & $-$ & $-$ & $-$ & \\
\hline\hline		
\end{tabular}
\end{center}

Similar to solutions \textbf{V} with an event horizon of $r_{H}=0.2$, configurations \textbf{VI} also possess a system of two unstable light rings and one ergoregion for each of the Gaussian curvatures of the target space $\kappa$. The main distinguishing feature here is that the Gaussian curvature has a negligible effect on the radii of the two light rings and the equatorial domain of the ergoregion. Specifically, the ergoregions extend from the event horizon and reach distances, $R_{ER}\in[0, \, 0.1968]$, $R_{ER}\in[0, \, 0.1972]$, and $R_{ER}\in[0, \, 0.1976]$, corresponding to negative, zero, and positive $\kappa$, respectively. Additionally, within the ergoregions, a prograde rotating light ring exists. With an increase in the curvature of the target space, the radii of those light rings decrease insignificantly, estimated as $R_{LR}=0.1599$, $R_{LR}=0.1598$, and $R_{EL}=0.1597$, for negative, zero, and positive $\kappa$, respectively. Moreover, the outer light rings, situated beyond the ergoregions, rotate retrograde relative to the black hole rotation and expand with increasing Gaussian curvature: $R_{LR}=0.4306$, $R_{LR}=0.4310$, and $R_{LR}=0.4314$, for negative, zero, and positive $\kappa$, respectively. Concurrently, as exposed in Fig. \ref{fig:8}, despite the different Gaussian curvature, the shadows exhibit an O-like shape, and due to the almost constant fraction of the horizon mass to the ADM mass, their area appears the same to an observer from the equatorial plane.

\section{Conclusion}

In this work we consider the shadow cast by rotating hairy black holes coupled to two non-trivial time-periodic scalar fields. The scalar fields may be viewed as coordinates in a nonphysical Riemannian space possessing a constant curvature. Thus, the black hole solutions can be divided into three classes according to the sign of its Gaussian curvature. Each class encompasses a wide range of solutions with different phenomenology depending on the amount of scalar hair, which can be measured by a normalized charge $q\in(1,0)$, representing the ratio between the scalar field Noether charge and the black hole spin. Solutions with high values of the normalized scalar charge approaching $q=1$ possess similar properties as the soliton-like solutions of the Einstein-scalar field equations representing boson stars. On the other hand, for small $q\approx 0$ the back-reaction of the scalar fields on the spacetime geometry vanishes and the solutions reduce to scalar clouds. Near this limit, the hairy black holes interact weakly with the scalar fields and resemble in their properties the Kerr black hole.

In our analysis, we examine the influence of the Gaussian curvature and the normalized charge $q$ on the properties of the hairy black hole shadow. For that purpose, we consider configurations with several representative values of $q$ and study the corresponding solutions for positive, negative and zero curvature of the target space. The shadows of the selected solutions are constructed numerically using a ray-tracing procedure.  We observe the following systematical behavior. The shadows for extremely high values of the normalized charge ($q>0.997$ for the presented models) exhibit large regions of chaotically scattering geodesics for all considered Gaussian curvatures. This leads to the formation of multiple disconnected shadow images. Increasing the Gaussian curvature, the chaotic behavior becomes milder producing more coherent shadow images. The chaotic behavior further reduces if we consider lower normalized charges. The shadows for charges in the range $q\sim 0.96-0.994$ become more compact possessing fewer disconnected components and a clearly defined dominant component. Positive Gaussian curvatures lead again to more regular shadow images compared to negative ones, as well as a larger size of the central dark region. Conversely, for negative Gaussian curvatures, the shadow size decreases substantially and the chaotic region dominates. Finally, for values of the normalized charge around $q=0.85$ the chaotic behavior of the scattering geodesics becomes negligible for all the Gaussian curvatures. For lower $q$ the shadow of the hairy black holes is represented by a single oval component and resembles qualitatively the Kerr black hole while its size is influenced relatively weakly by the Gaussian curvature. 

The presence of chaotic regions and the shadow shape significantly influence the optical appearance of accretion disks around hairy black holes. Exploring this problem is ongoing work, but we can offer some qualitative insights. What matters most when calculating the actual image produced by a given accretion disk around the black hole is the presence of a central compact shadow. Having extensive chaotic regions, like those observed very close to the boson star limit with $q$ approaching $1$, would likely not produce significant or any dark region of the image, which practically contradicts present observations. Thus, models with very large $q$, or models with hefty scalar hair, can be excluded by the Event Horizon Telescope observations. Large positive Gaussian curvature can potentially relieve this problem because it increases the size of the central dark shadow and decreases the chaotic regions close to the boson star limit. In the present paper, we have considered only moderately high curvatures since the models calculated in \cite{collodel2020rotating} were taken as a background. It will be interesting to investigate whether this pattern remains with a further increase in Gaussian curvature and whether one can extend the dark region of the shadow, as supported by current observations, even close to the boson star limit ($q=1$).

\section*{Acknowledgements}
This study is financed by the European Union-NextGenerationEU, through the National Recovery and Resilience Plan of the Republic of Bulgaria, project No. BG-RRP-2.004-0008-C01. DD acknowledges financial support via an Emmy Noether Research Group funded by the German Research Foundation (DFG) under grant no. DO 1771/1-1.

\begin{appendix}

\section{Physical quantities of selected solutions}\label{A1}
Below we present in detail the characteristics of the solutions used to produce the black hole images. They are taken from \cite{collodel2020rotating} where sequences of constant horizon radii were constructed for various Gaussian curvatures $\kappa$.

\begin{table}[h!]
    \centering
    \setlength{\tabcolsep}{6.6pt} 
    \renewcommand{\arraystretch}{1.3} 
	\begin{tabular}{lcccccccccc}
        \hline
        \hline
        Label & $\omega_{s}/\mu$ & $M\mu$    & $J \mu^2$ & $M_{BH}\mu$ & $J_{BH}\mu^2$ & $M_{\psi}\mu$ & $ J_{\psi}\mu^2$ & $ J_{\psi}/J$ & $J/M^2$  & $J_{BH}/M_{BH}^2$ \\ 
        
        \hline
        
        \textbf{I}$^{\,-5}_{\,0.01}$       & 0.7398         & 0.6902  & 0.4734  & 0.0056    & 0.0013      & 0.6778      & 0.4721         & 0.9972      & 0.9938  & 42.544  \\         
        \textbf{I}$^{\,0}_{\,0.01}$  & 0.8353         & 0.6482  & 0.4068  & 0.0049    & 0.0011      & 0.6434      & 0.4057         & 0.9973      & 0.9680  & 46.444  \\ 
        \textbf{I}$^{\,5}_{\,0.01}$  & 0.8219         & 0.7425  & 0.5041 & 0.0049  & 0.0015    & 0.7469      & 0.5026      & 0.9971 &   0.9144    &  60.307  \\   
        
        \hline
        
        \textbf{II}$^{\,-5}_{\,0.01}$       & 0.6074 & 0.8905 & 0.7813 & 0.0039 & 0.0001 & 0.8576 & 0.7812 & 0.9999 & 0.9853 & 6.7133  \\ 
        \textbf{II}$^{\,0}_{\,0.01}$  & 0.6792 & 0.8820 & 0.7259 & 0.0035 & 0.0001 & 0.8785 & 0.7258 & 0.9999 & 0.9331 & 5.8772  \\ 
        \textbf{II}$^{\,5}_{\,0.01}$  & 0.7316 & 0.8832 & 0.6890 & 0.0031 & 0.0002 & 0.8984 & 0.6888 & 0.9997 & 0.8833 & 11.727  \\   
        
        \hline
        
        \textbf{III}$^{\,-5}_{\,0.05}$       & 0.6537 & 0.8865 & 0.7567 & 0.0256 & 0.0042 & 0.8244 & 0.7525 & 0.9944 & 0.9629 & 6.3547 \\ 
        \textbf{III}$^{\,0}_{\,0.05}$  & 0.7064 & 0.9082 & 0.8329 & 0.0252 & 0.0045 & 0.9415 & 0.8283 & 0.9945 & 0.8915 & 7.1310 \\  
        \textbf{III}$^{\,5}_{\,0.05}$  & 0.7023 & 1.1003 & 0.9812 & 0.0254 & 0.0052 & 1.1078 & 0.9760 & 0.9947 & 0.8105 & 8.0147  \\ 
        
        \hline
        
        \textbf{IV}$^{\,-5}_{\,0.1}$       & 0.7138 & 0.8738 & 0.7122 & 0.0780 & 0.0248 & 0.7639 & 0.6875 & 0.9653 & 0.9327 & 4.0651 \\ 
        \textbf{IV}$^{\,0}_{\,0.1}$  & 0.7385 & 1.0004 & 0.8535 & 0.0870 & 0.0302 & 0.9134 & 0.8232 & 0.9645 & 0.8528 & 3.9982 \\   
        \textbf{IV}$^{\,5}_{\,0.1}$  & 0.7487 & 1.1610 & 1.0394 & 0.0904 & 0.0322 & 1.1001 & 1.0071 & 0.9689 & 0.7711 & 3.9335 \\ 
        
        \hline
        
        \textbf{V}$^{\,-5}_{\,0.2}$        & 0.8835 & 0.8474 & 0.6478 & 0.2706 & 0.0976 & 0.5709 & 0.5499 & 0.8489 & 0.9022 & 1.3339 \\ 
        \textbf{V}$^{\,0}_{\,0.2}$   & 0.8955 & 0.8787 & 0.6790 & 0.2796 & 0.1012 & 0.5990 & 0.5776 & 0.8508 & 0.8793 & 1.2946 \\   
        \textbf{V}$^{\,5}_{\,0.2}$   & 0.9078 & 0.8847 & 0.6814 & 0.2921 & 0.1065 & 0.5951 & 0.5746 & 0.8432 & 0.8707 & 1.2486 \\ 
              
        \hline
        
        \textbf{VI}$^{\,-5}_{\,0.3}$        & 0.9880 & 0.3190 & 0.1252 & 0.2373 & 0.0442 & 0.0817 & 0.0810 & 0.6469 & 1.2300 & 0.7848 \\ 
        \textbf{VI}$^{\,0}_{\,0.3}$   &  0.9880 & 0.3200 & 0.1260 & 0.2377 & 0.0444 & 0.0823 & 0.0816 & 0.6478 & 1.2304 & 0.7856 \\   
        \textbf{VI}$^{\,5}_{\,0.3}$   &  0.9880 & 0.3210 & 0.1268 & 0.2380 & 0.0446 & 0.0830 & 0.0823 & 0.6486 & 1.2306 & 0.7865 \\
        
        \hline
        
        \textbf{VII}$^{\,-5}_{\,0.05}$  & 0.6485 & 0.9157 & 0.8048 & 0.0243 & 0.0030 & 0.8493 & 0.8017 & 0.9962 & 0.9598 & 5.0973 \\        
        \hline
        \hline
    \end{tabular}    
\end{table}
\begin{table}
    \centering
    \setlength{\tabcolsep}{6.6pt} 
    \renewcommand{\arraystretch}{1.3} 
    \begin{tabular}{lcccccccccc}
 
    \end{tabular}
    \caption{\label{tab_7}Physical quantities of hairy black hole solutions as depicted in Fig. \ref{fig:kappa}, corresponding to different Gaussian curvatures of the target space, $\kappa=\{-5, \, 0, \, 5\}$. Each configuration is identified by \textbf{X}$^{\,u}_{\,v}$, where the symbol \textbf{X} represents the configuration number, superscript $u$ represents the value of the Gaussian curvature $\kappa$, and the subscript $v$ denotes the value of the black hole horizon $r_{H}$. In a single group of solutions the horizon radii $r_H$ is constant and the normalized charge $q = \frac{J_{\psi}}{J}$ (represented in the 9th column) is adjusted to be nearly constant.}  
\end{table}

\end{appendix}

\end{document}